\documentclass[a4paper,fleqn,usenatbib]{mnras}
\usepackage{lscape,graphicx}
\usepackage{rotating}
\usepackage{color}
\usepackage{threeparttable}
\usepackage{amsmath,amssymb}

\usepackage{color}

\usepackage[normalem]{ulem}  
\usepackage{color,soul}

\newcommand{\old}[1]{}

\newcommand{\avk}[1]{{#1}}
\newcommand{\aad}[1]{{#1}}
\newcommand{\yus}[1]{{#1}}

\def\an{Astron. Nachr.}

\def\asec{\ifmmode ^{\prime\prime}\else$^{\prime\prime}$\fi}

\def\Msun{M$_{\odot}$}

\def\it{\sl}
\def\degs{\ifmmode ^{\circ}\else$^{\circ}$\fi}
\def\amin{\ifmmode ^{\prime}\else$^{\prime}$\fi}
\def\asec{\ifmmode ^{\prime\prime}\else$^{\prime\prime}$\fi}
\def\fdg{\hbox{$.\!\!^\circ$}}          
\def\farcs{\hbox{$.\!\!^{\prime\prime}$}}  
\def\h{$^{\rm h}$}
\def\m{$^{\rm m}$}

\def\degs{\ifmmode ^{\circ}\else$^{\circ}$\fi}
\def\amin{\ifmmode ^{\prime}\else$^{\prime}$\fi}

\def\eqalign#1{\null\,\vcenter{\openup1\jot \m@th
   \ialign{\strut\hfil$\displaystyle{##}$&$\displaystyle{{}##}$\hfil
   \crcr#1\crcr}}\,}
\def\j0633{J0633}
\def\xmm{\textit{XMM-Newton}}
\def\chan{\textit{Chandra}}
\def\fermi{\textit{Fermi}}

\title[\aad{\xmm\ observations of PSR J0633+0632}]{\aad{\xmm\ observations of a gamma-ray pulsar J0633+0632: pulsations, cooling and large-scale emission}}
\author[A.~Danilenko, A.~Karpova, D.~Ofengeim, Yu.~Shibanov and D.~Zyuzin]
{A.~Danilenko,\thanks{E-mail: danila@astro.ioffe.ru}
A.~Karpova, 
D.~Ofengeim, Yu.~Shibanov
and
D.~Zyuzin
\\
Ioffe Institute, Politekhnicheskaya 26, St. Petersburg, 194021, Russia}

\date{Accepted XXX. Received YYY; in original form ZZZ}
\pubyear{2020}

\begin{document}

\label{firstpage}
\pagerange{\pageref{firstpage}--\pageref{lastpage}}
\maketitle

\begin{abstract}
We report results of \xmm\ observations of a $\gamma$-ray pulsar J0633+0632 and its wind nebula. \aad{We reveal, for the first time, pulsations of the pulsar X-ray emission with a single sinusoidal pulse-profile and a pulsed fraction of $23 \, \pm \, 6$ per cent in the 0.3--2 keV band.} We confirm previous \chan\ findings that the pulsar X-ray spectrum consists of thermal and non-thermal components. However, we do not find the absorption feature that was previously detected at about 0.8 keV. Thanks to the greater sensitivity of \xmm, we get stronger constraints on spectral model parameters compared to previous studies. The thermal component can be equally well described by either blackbody or neutron star atmosphere models, implying that this emission is coming from either hot pulsar polar caps with a temperature of about 120 eV or from the colder bulk of the neutron star surface with a temperature of about 50 eV. In the latter case, the pulsar appears to be one of the coolest among other neutron stars of similar ages with estimated surface temperatures. We discuss cooling scenarios relevant to this neutron star. Using an interstellar absorption--distance relation, we also constrain the distance to the pulsar to the range of 0.7--2 kpc. Besides the pulsar and its compact nebula, we detect regions of weak large-scale diffuse non-thermal emission in the pulsar field and discuss their possible nature.
\end{abstract}


\begin{figure*}
\begin{minipage}[h]{0.495\linewidth}
\center{\includegraphics[scale=0.51,clip]{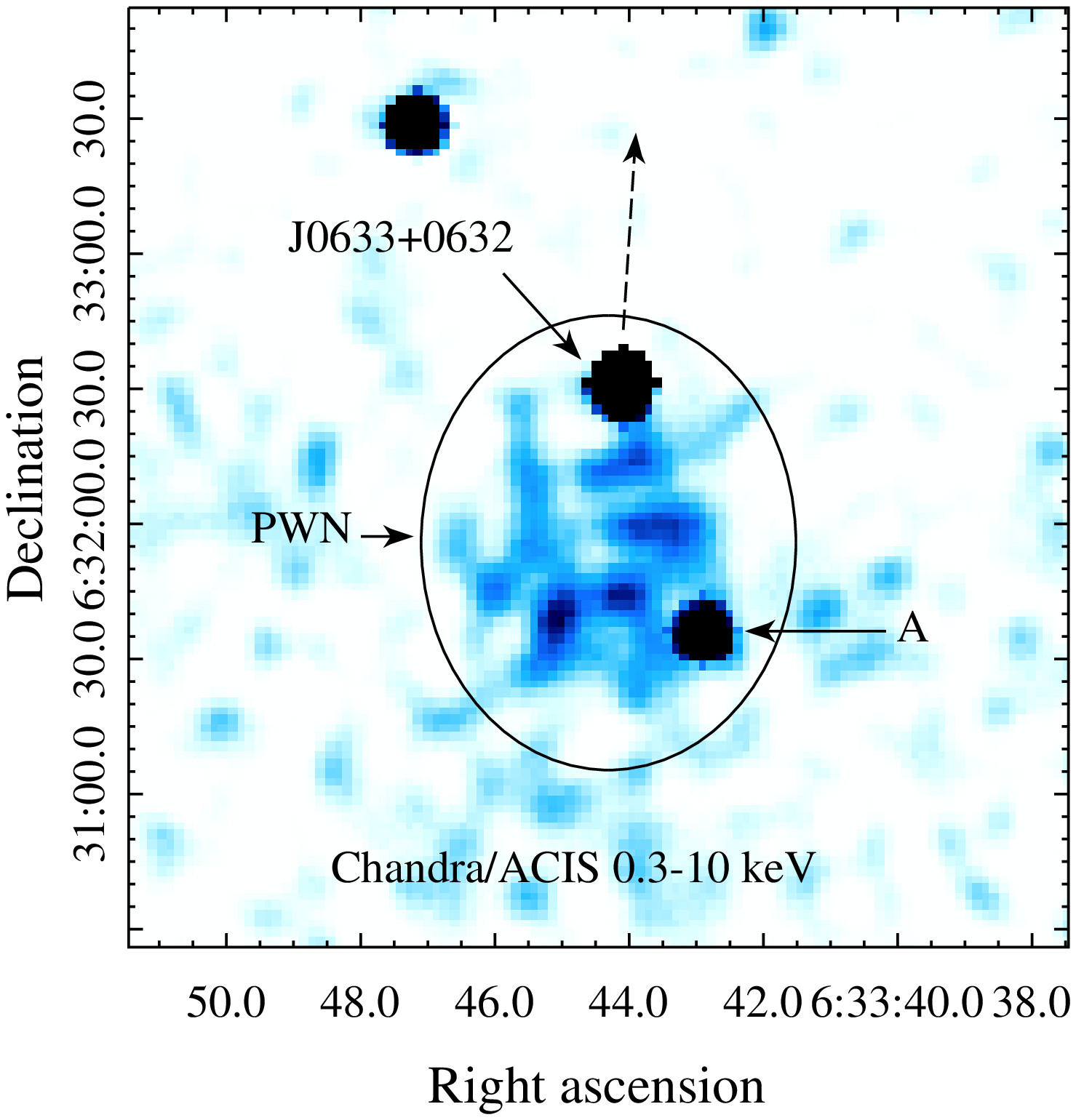}}
\end{minipage}
\begin{minipage}[h]{0.495\linewidth}
\center{\includegraphics[scale=0.5,clip]{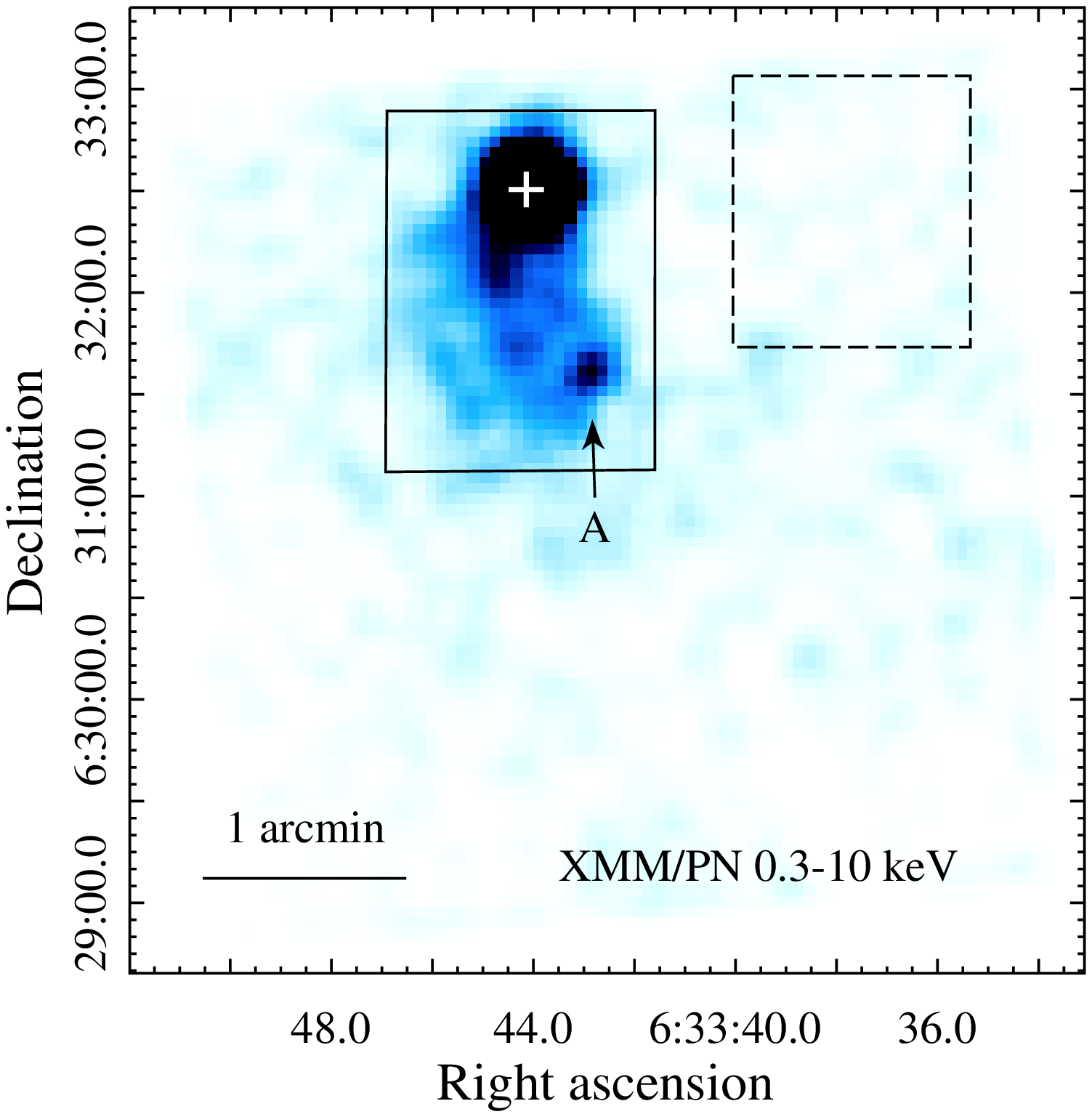}}
\end{minipage}
\caption{{\sl Left}: 3.5 $\times$ 3.5 arcmin \chan\ image of the \j0633\ vicinity in the 0.3--10 keV band. \avk{\j0633\ and an unrelated source `A' are marked. The compact PWN is enclosed by the ellipse.} The possible direction of the pulsar proper motion is shown by the dashed arrow. {\sl Right}: FOV of the \xmm\ EPIC-pn camera in the Small Window mode (0.3--10 keV). The \j0633\ position is shown by the `+' symbol. Source `A' is also marked. The solid and dashed boxes enclose regions used to extract the PWN and background spectra.}
\label{fig:chandra}
\end{figure*}

\begin{figure*}
\begin{minipage}[h]{0.495\linewidth}
\center{\includegraphics[scale=0.52,clip]{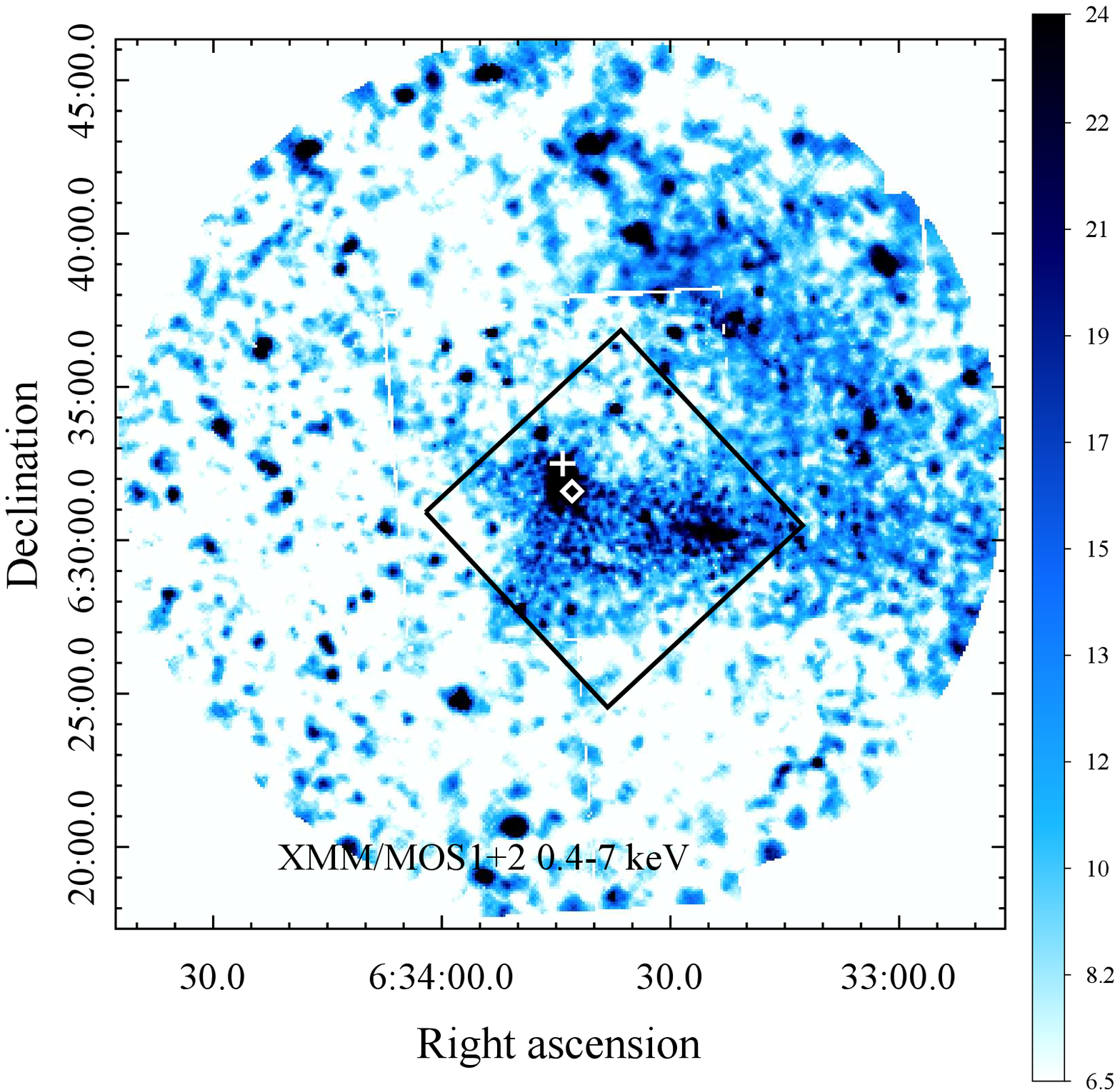}}
\end{minipage}
\begin{minipage}[h]{0.495\linewidth}
\center{\includegraphics[scale=0.52,clip]{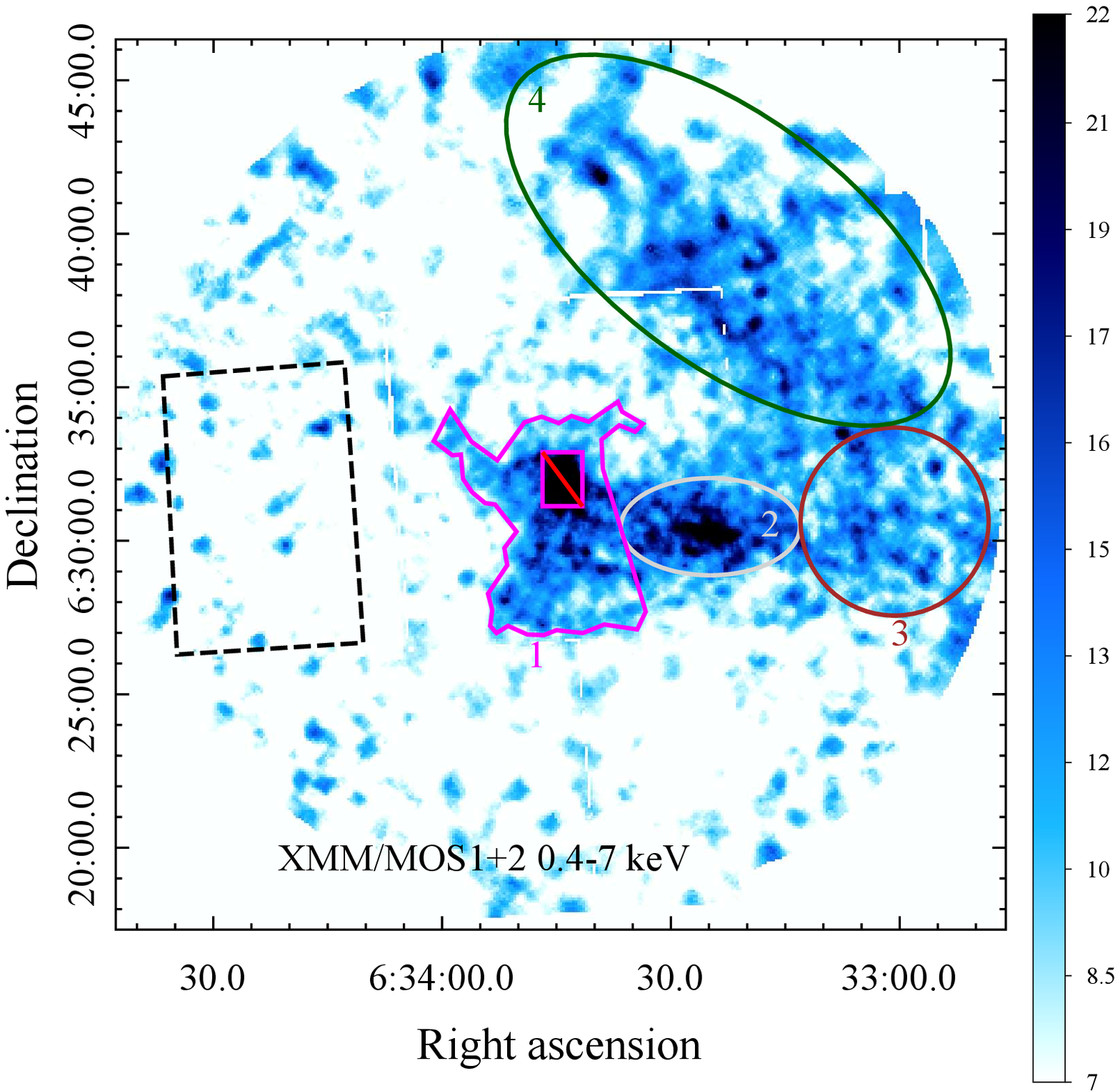}}
\end{minipage}
\begin{minipage}[h]{0.495\linewidth}
\center{\includegraphics[scale=0.52,clip]{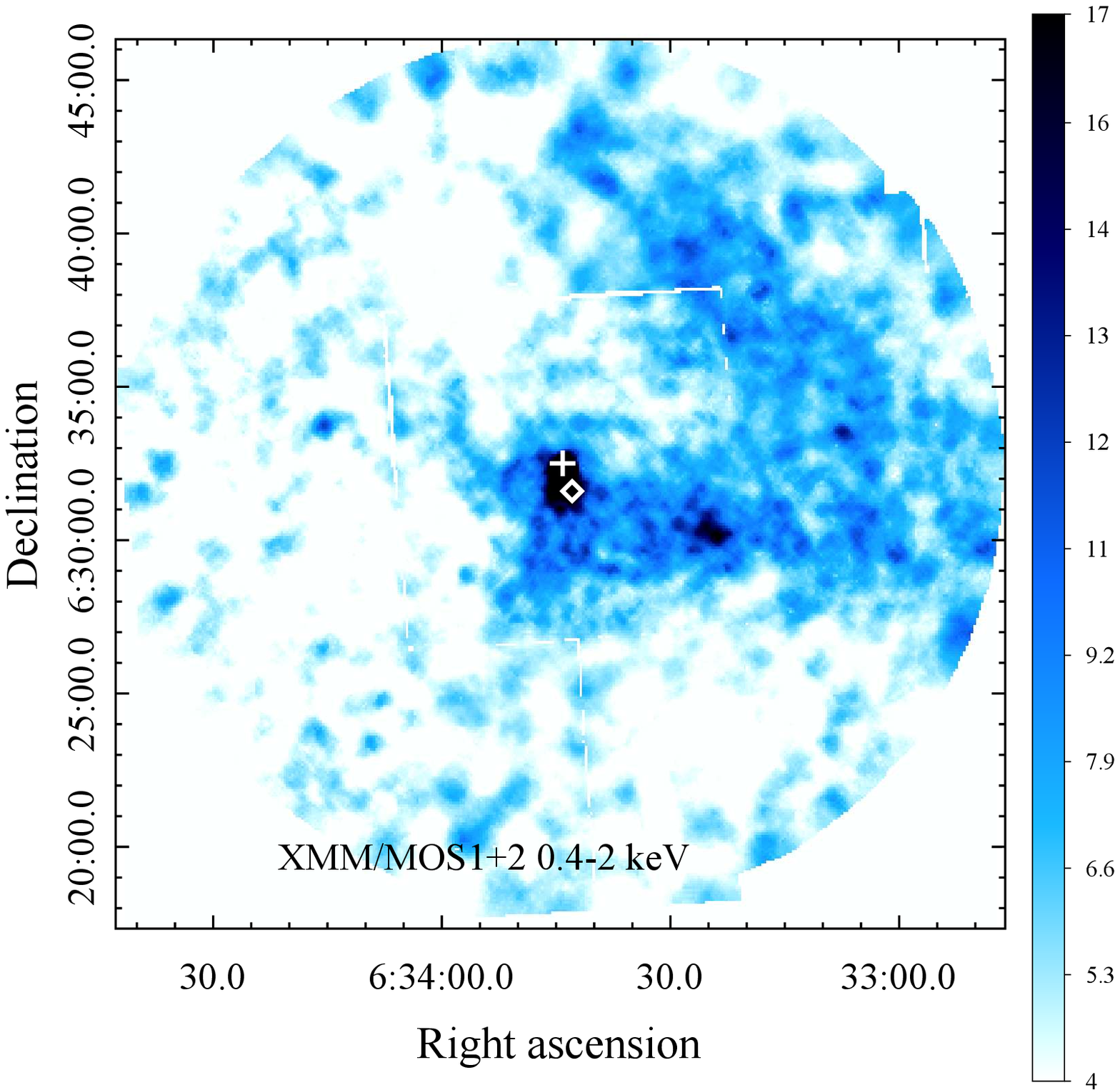}}
\end{minipage}
\begin{minipage}[h]{0.495\linewidth}
\center{\includegraphics[scale=0.52,clip]{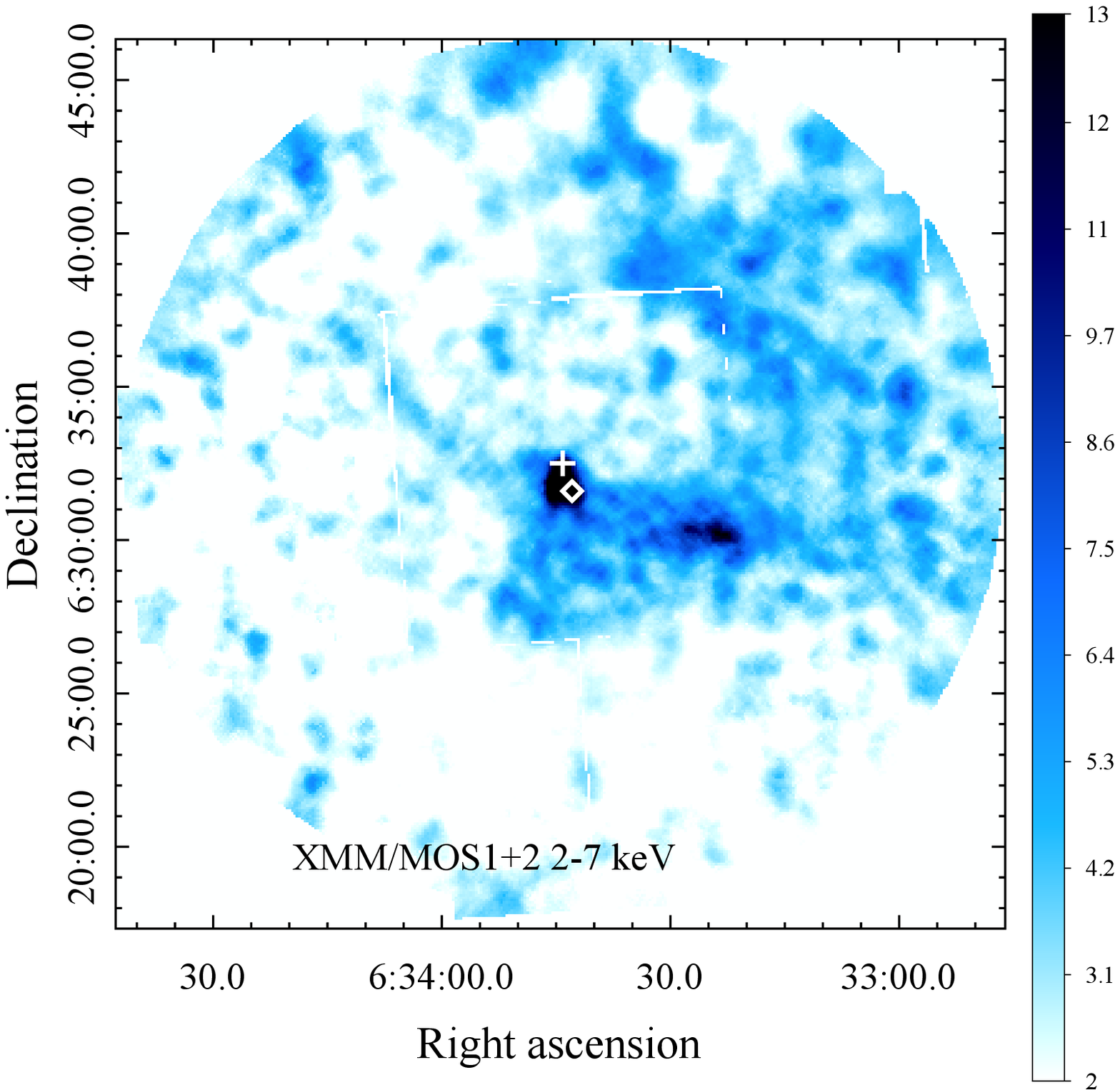}}
\end{minipage}
\caption{\avk{{\sl Top-left}: combined \xmm\ MOS1 and MOS2 exposure-corrected QPB-subtracted image of the \j0633\ field in 0.4--7 keV band. The positions of \j0633\ and the `A' source are shown by the `+' and `$\Diamond$' symbols, respectively. The solid box in the {\sl top-left} indicates the \textit{Chandra}/ACIS-S3 chip FOV. {\sl Top-right}: the same combined image where we removed all point sources except of \j0633\ and `A'.} Regions used for spectral analysis of the diffuse emission are shown and numbered. The bright PWN region shown in the right-hand panel of Fig.~\ref{fig:chandra} was excluded from the region 1 in the analysis. The dashed box was used for the background. \avk{{\sl Bottom}: the same as in the {\sl top-right} but in the soft and hard photon energy bands, as shown in the panels.} The intensity is given in counts s$^{-1}$ deg$^{-2}$.}
\label{fig:j0633comb}
\end{figure*}

\section{Introduction}
\label{sec:intro}

More than 200 pulsars have been detected in $\gamma$-rays with the \textit{Fermi} observatory.\footnote{\href{https://confluence.slac.stanford.edu/display/GLAMCOG/Public+List+of+LAT-Detected+Gamma-Ray+Pulsars}{All the \textit{Fermi} detected pulsars.}} A significant number of them are radio-quiet, which can be partially explained by unfavourable orientations of pulsar radio beams. Due to the intrinsic faintness of pulsars in the optical band, any additional information on the radio-quiet \textit{Fermi} pulsars, including distances to them, can be obtained only from observations in X-rays. X-ray observations are also crucial for the study of non-thermal and thermal emission components from pulsar magnetospheres and surfaces. In addition, X-ray observations can reveal a parent supernova remnant (SNR) and/or a pulsar wind nebula (PWN), studies of which can help to constrain the age and transverse velocity of the corresponding pulsar as well as properties of its wind and environment. \aad{Furthermore, analysing the morphology of some PWNe, namely torus-like PWNe like Crab and Vela, one can constrain the inclination of the pulsar's rotational axis to the observer's line of sight \citep{ng2004,ng2008}.} 

A radio-quiet pulsar J0633+0632 (hereafter J0633) was discovered with \textit{Fermi} by \citet{abdo2009}. It has a period $P$~= 297.4~ms, a characteristic age $t_c$~= 59.2~kyr, a spin-down luminosity $\dot{E}$~= $1.2\times10^{35}$~erg~s$^{-1}$ and a dipole surface magnetic field $B$~= $4.9\times10^{12}$~G \citep{abdo2013ApJS}. The first 20-ks \chan\ observations allowed \citet{ray2011} to identify \j0633\ in X-rays and to reveal a faint PWN adjusted to the pulsar. The \chan\ image obtained with the Advanced CCD Imaging Spectrometer (ACIS) is shown in the left panel of Fig.~\ref{fig:chandra} where \j0633, its PWN and an unrelated point-like source `A' are marked \aad{\citep[see][for a description of the data reduction]{danilenko2015}}. The X-ray spectrum of \j0633\ consists of thermal and non-thermal components \citep{ray2011,danilenko2015}. The later is fitted by a power law (PL) while the former can be equally well described by either the blackbody model or the model of \aad{a neutron star (NS) magnetized atmosphere \citep{pavlov1995,ho2008}}.

\citet{danilenko2015} found a signature of an absorption feature at $\approx 0.8$ keV in the \chan\ spectrum of the pulsar. They suggested that it might be the cyclotron line created in the strong magnetic field of the NS though other origins are also possible. Among known isolated NSs, X-ray absorption lines have been  reported only for a few exotic objects, including compact central objects in SNRs, X-ray dim isolated NSs and magnetars \citep[see e.g.][for references]{danilenko2015}. \old{No lines have been found}\aad{Concerning rotation-powered pulsars (RPPs), the most numerous subclass of isolated NSs, there are only PSR J1740$+$1000 \citep{kargaltsev2012} and possibly PSR J0659+1414 \citep{arumugasamy2018} whose spectra show absorption features}. \j0633\ could thus be the third \aad{such RPP}\old{showing an absorption feature}. However, low count statistics of the \chan\ data does not allow one to confidently resolve its \aad{line} profile.

Analysing the interstellar absorption towards \j0633, \citet {danilenko2015} constrained the distance to the pulsar within the range of 1--4 kpc. They also noted that the elongation  of the PWN southwards of the pulsar is likely caused by the pulsar proper motion in the opposite direction. The presumed proper motion  direction is shown by the dashed arrow in Fig.~\ref{fig:chandra} (left-hand panel). Its expected value was estimated to be $\approx$~80~mas~yr$^{-1}$, which corresponds to a transverse velocity of 380$D_{\rm 1kpc}$~km~s$^{-1}$, where $D_{\rm 1kpc}$ is the distance in units of 1 kpc. Taking this, a possible birthplace of \j0633\ was suggested to be in the Rosette nebula, which is a 50-Myr-old active star-forming region.

Searching for \j0633\ in the optical has been performed with the Gran Telescopio Canarias \citep{mignani2016}. No pulsar optical counterpart was detected down to $\approx$27.3 magnitude in the $g$ band. The limit is consistent with the  extrapolation of the X-ray PL spectral component to the optical range. The High Altitude Water Cherenkov (HAWC) collaboration recently reported on the possible detection of a TeV halo around \j0633, HAWC J0635+070, extending by about 0\fdg 65 and recalling the TeV halo around the Geminga pulsar \citep{brisbos2018}.

To further study \j0633 in X-rays, we performed deeper observations\footnote{ObsID 0764020101, PI Danilenko} with \xmm.\old{Preliminary results were briefly reported in a conference presentation \citep{2017JPhCSkarpan}. \xmm\ detected the pulsing X-ray emission from the pulsar and also revealed a large-scale diffuse emission around it.} Here we present \aad{a description of} the \old{explicit \xmm}data analysis and \old{final}results. The paper is organized as follows. The data and imaging analysis are described in Section~\ref{sec:data}. Timing and spectral analysis of \j0633\ are presented in Sections~\ref{sec:timing} and \ref{sec:j0633spec}, respectively. \aad{In Section~\ref{sec:diffuse}, we analyse the large-scale diffuse emission revealed around the pulsar by \textit{XMM-Newton}.} We discuss the results in Section~\ref{sec:discussion} and give a short summary in Section~\ref{sec:sum}.

\section{The data and imaging analysis}
\label{sec:data}

The \j0633\ field was observed with the European Photon Imaging Camera (EPIC)\footnote{\url{https://www.cosmos.esa.int/web/xmm-newton/technical-details-epic}} onboard \xmm\ on 2016 March 31 (MJD 57478), with a total exposure time of 93 ks. Two metal oxide semi-conductor (MOS) CCD arrays were in the Full Frame mode with the medium filter setting while the pn-CCD  detector (EPIC-pn) was in the small window mode with the thin filter enabling timing data analysis with $\approx 6$ ms temporal resolution. We used the {\sc xmm-sas} v.16.0.0 software for the data analysis.

We exclude periods of high background activity using the {\sc espfilt} tool. This results in clean exposure times of 51.8, 63.6 and 33.0 ks for the MOS1, MOS2 and pn cameras, respectively.

The EPIC-pn field-of-view (FOV) is shown in the right-hand panel of Fig.~\ref{fig:chandra}. As seen, the pulsar and its PWN, previously revealed with \chan, are firmly detected with \xmm. The \chan\ position of \j0633\ (see Table~\ref{tab:ephem}) is marked by the `+' symbol. The image appears to be blurred, as compared to the \chan\ image, due to the lower spatial resolution of \xmm. Nevertheless, an unrelated point-like background source `A', with coordinates RA~=~6\h33\m42\fs902 and Dec.~=~+6\degs31\amin36\farcs16, obtained with a {\sc ciao} tool {\sc wavdetect} from the \chan\ data, is clearly resolved from the PWN in the EPIC-pn image.

Due to the mode selected, the EPIC-pn data allow us to image only the nearest vicinity of the pulsar, constrained by a small FOV of $\approx$4~$\times$~4~arcmin. We used MOS1 and MOS2 data and the \textit{XMM-Newton} Extended Source Analysis Software \citep[{\sc xmm-esas};][]{cookbook} to construct much larger images with a FOV of $\approx$30~$\times$~30~arcmin.\footnote{Note that MOS1 CCDs 3 and 6 were damaged due to micrometeorite strikes and thus switched off \citep{cookbook}.} We created these images and respective exposure maps using the {\sc mos-spectra} tool. The quiescent particle background (QPB) images were generated by the {\sc mos\_back} task and then subtracted. We adaptively smoothed the MOS1$+$MOS2 QPB-subtracted and exposure-corrected image applying the {\sc adapt} tool and accumulating 50 counts for the smoothing kernel. The resulting image, in the 0.4--7.0 keV energy band, is presented in the top left-hand panel of Fig.~\ref{fig:j0633comb}.

Besides the pulsar and its compact PWN seen with \chan\ and \xmm/EPIC-pn, this image also reveals a fainter extended emission at larger scales. A relatively bright emission clump located west of the compact PWN and a long extended structure in the north-western part of the image, which apparently is not related to the pulsar, are particularly interesting. To better investigate them, we also  created images in the 0.4--7, 0.4--2 and 2--7 keV bands where point-like sources were removed and the respective holes were refilled utilizing the {\sc ciao dmfilth} task and pixel values from surrounding background regions. We did not exclude \j0633\ and the `A' source since their removal leads to some distortion of the compact PWN shape. MOS1+MOS2 images were then adaptively smoothed accumulating 100 counts for the smoothing kernel. They are presented in the top-right and bottom panels of Fig.~\ref{fig:j0633comb}. One can see that morphology of the extended emission is roughly the same in the soft and hard bands, although the emission intensity appears to be higher at lower energies.

\section{Timing analysis}
\label{sec:timing}

We used the EPIC-pn data to search for pulsations from \j0633. To obtain maximal sensitivity for a pulsing component, we did not filter the event list for flaring background and use events in the 0.3--10 keV range extracted from a 15 arcsec-radius aperture centred at the \textit{Chandra} position of the pulsar. This resulted in the total event number of 1717. We then corrected the event times of arrival (ToA) to the Solar system barycentre using the {\sc sas} task {\sc barycen}, the \j0633\ \textit{Chandra} coordinates obtained by \citet{ray2011} (see Table~\ref{tab:ephem}) and the Solar system ephemeris DE~405.

\begin{figure}
  \begin{center}
    \includegraphics[width=\columnwidth]{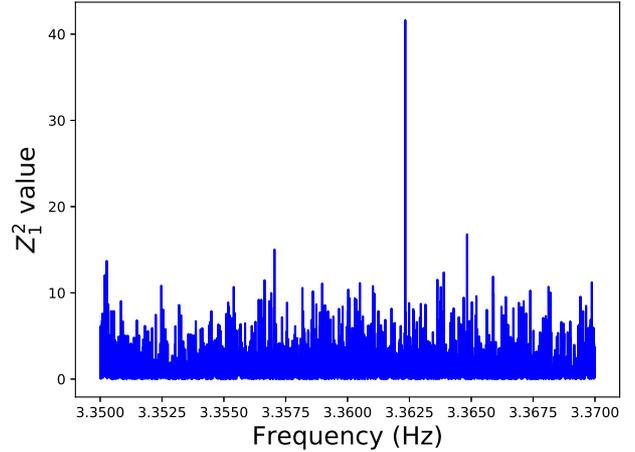}
  \end{center}
  \caption{$Z^2_1$-test periodogram for \textit{XMM-Newton} observations of \j0633 using only events with energies in the range of 0.3--10~keV. The event list was not filtered for flaring background, to obtain maximal sensitivity for the pulsing component.}
  \label{fig:ztest}
\end{figure}

As a first step, we examined the $Z^2_1$-test periodogram \citep{ztest} in the frequency range of 3.35--3.37 Hz, enclosing the pulsar rotation frequency known from \textit{Fermi} data (Table~\ref{tab:ephem}). It shows a pronounced peak at the frequency of $\approx3.362332$~Hz with $Z^2_{\rm 1}\approx40$, which corresponds to the pulsation detection significance of $\approx4.8\sigma$ (Fig.~\ref{fig:ztest}).\footnote{\avk{The corresponding frequency uncertainty, calculated using the formula from \citet{chang2012}, is 1.3 $\mu$Hz.}} The X-ray pulsation frequency is consistent with the $\gamma$-ray one of 3.362332235(3)~Hz, which is adjusted to the epoch of the \textit{XMM-Newton} observations (MJD 57478) using the \textit{Fermi} timing results from Table~\ref{tab:ephem}.

To crosscheck this result and to compute the frequency uncertainty, we applied the Gregory-Loredo Bayesian method for the analysis of periodic signals \citep{gregory&loredo1992} \avk{and used pn data cleaned from the flaring background.} The method considers a number of step-wise profiles each consisting of a specific number of steps $m$, $m=1 ... m_{\rm max}$. During the analysis, we folded ToA with each $m$-step model for any trial pair of frequency $\nu$ and phase $\phi$. The method then applies the Bayesian theory to compute a probability $p$ in favour of any periodic model ($m>1$) over the constant model ($m=1$):
\begin{equation}
 \label{eq:signalsignificance}
  p(m>1|{\rm ToA}) = \dfrac{\sum_{m=2}^{m_{\rm max}}O_{m1}}{1+\sum_{m=2}^{m_{\rm max}}O_{m1}}.
\end{equation}
In equation~(\ref{eq:signalsignificance}), $O_{m1}$ is the odds ratio in favour of the $m$-step periodic model. It is inversely proportional to the number of ways a given distribution $\{n_{i}\}$, $i=1 ... m$, of $N_{\rm ToA}$ times of arrival over $m$ period bins could have arisen by chance, or the so-called multiplicity:
\begin{equation}
  \label{eq:multiplicity}
    W_{m}(\nu,\phi) = \dfrac{N_{\rm ToA}!}{\prod_{i=1}^{m}n_{i}!}.
\end{equation}
To sample the probability density $-\ln{W_m(\nu,\phi)}$, we applied the Metropolis-Hastings (MH) Markov chain Monte Carlo (MCMC) method \citep{metro1953}.

\begin{table}
  \renewcommand{\arraystretch}{1.2}
  \caption{The \j0633 timing model.$^{\dag}$}
  \label{tab:ephem}
  \begin{center}
    \begin{tabular}{lc}
      \hline
      R.A. (J2000)$^\ddag$                                & 06\h33\m44\fs142 \\
      Dec. (J2000)$^\ddag$                                & +06\degs32\amin30\farcs40 \\
      Rotations frequency $f$, Hz                         & 3.3624817298(6)$^\S$
      \\
      Frequency derivative $\dot{f}$, Hz s$^{-1}$         & $-$8.9983(2)$\times$10$^{-13}$ \\
      Frequency second derivative $\ddot{f}$, Hz s$^{-2}$ & 7.2$\times$10$^{-25}$ \\
      Epoch of frequency, MJD                             & 55555 \\
      Valid MJD range                                     & 54686.15--56583.16 \\
      Solar system ephemeris model                        & DE405 \\
      Time system                                         & TDB \\
      \hline
    \end{tabular}
  \end{center}
  \begin{tablenotes}
    \item $^{\dag}$ Obtained from the LAT Gamma-ray Pulsar Timing Models page \citep{kerr2015} available at \url{http://www.slac.stanford.edu/~kerrm/fermi_pulsar_timing/}.
    \item $^{\ddag}$ The position is obtained from the \chan\ data \citep{ray2011}.
    \item $^\S$ Hereafter, the numbers in parentheses denote errors relating to the last significant
    digit quoted.
  \end{tablenotes}
\end{table}

\old{In Fig.~\ref{fig:odds}, we show $O_{m1}$ values versus the number of phase bins $m$. It is seen that}\aad{It turns out} $O_{m1}$ drops very quickly as $m$ increases. To compute the resulting odds ratio $Q_{\rm per}=\sum_{m=2}^{m_{\rm max}}O_{m1}$ in favor of the hypothesis that the signal is periodic, we thereby safely chose $m_{\rm max}=6$ since larger $m$ would not significantly contribute to $Q_{\rm per}$. The resulting $Q_{\rm per}\approx 2451$ corresponds to the 99.96 per cent probability (equation~\ref{eq:signalsignificance}) that the signal is periodic. \old{As an example, in Fig.~\ref{fig:freqpdf}, we show the marginal posterior probability density of the frequency obtained from the MCMC simulations for $m=5$.}\aad{MCMC simulations yield, for any $m$, a maximal-probability frequency of 3.362333(1) Hz, where the number in brackets is the uncertainty computed from the 68 per cent credible interval}. Within 1$\sigma$ uncertainty, it is consistent with the \textit{Fermi} value and with the results of the $Z^2_1$ test.

\begin{figure}
  \includegraphics[width=\columnwidth]{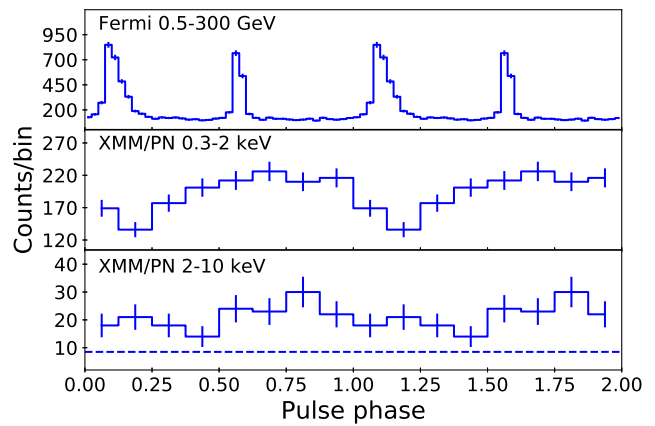}
  \caption{Phase-folded light curves of \j0633\ in different energy ranges, indicated in the panels.\old{(periods of flares were excluded from the data)} \aad{X-ray light curves were obtained from the data cleaned from flaring background}. The dashed line indicates the background level in the 2--10 keV band. In the 0.3--2 keV band, the background level ($\approx$~13 counts/phase bin) is not shown since it is significantly lower than the pulsar count rate.}
  \label{fig:xmmvsfermi}
\end{figure}

To calculate a zero rotational phase for EPIC-pn events, we used the \textit{Fermi} ephemeris and applied the {\sc photons} plug-in\footnote{\url{http://www.physics.mcgill.ca/~aarchiba/photons_plug.html}} for {\sc tempo2} package \avk{\citep*{hobbs2006mnras}}. Phase-folded \xmm\ light curves of \j0633\ in soft (0.3--2 keV) and hard (2--10 keV) bands are presented in the two bottom panels of Fig.~\ref{fig:xmmvsfermi}. We also show the $\gamma$-ray pulse profile obtained from \textit{Fermi} data using the {\sc tempo2} {\sc fermi} plug-in.\footnote{\avk{The J0633 \textit{Fermi} timing model, which was used to create the pulse profiles in Fig.~\ref{fig:xmmvsfermi}, is constructed for the MJD range (see Table \ref{tab:ephem}), which ends before \xmm\ observations. Since the extrapolated frequency is in agreement with that one found in the X-ray data, we just folded the pn light curve using this model. However, further refinement of the model potentially may lead to some change in the shift between the peaks of the X-ray and $\gamma$-ray profiles.}} \avk{For the latter, we downloaded data from the \fermi\ website\footnote{\url{https://fermi.gsfc.nasa.gov/cgi-bin/ssc/LAT/LATDataQuery.cgi}} and processed them using the \fermi\ Science tools (v10r0p5). We selected events from the 0\fdg8 radius aperture applying a SOURCE class events ({\sc evclass}=128) and a zenith angle of $<100$\degs. Good time intervals were generated assuming filtering criteria {\sc data\_qual} == 1 and {\sc lat\_config} == 1.}

The \avk{\j0633} pulsations are clearly seen in the soft \avk{X-ray} band, while they are only marginally resolved in the hard band. In contrast to the sharp double-peaked $\gamma$-ray pulse profile, with about 0.5 phase gap between two peaks, presumably produced by energetic particles \yus{accelerated near the equatorial current sheet, which emerges at and/or beyond the light cylinder of the pulsar \citep[see e.g.][and references therein]{2019kalapotharakos},} the soft X-ray profile is broad and sinusoidal, as expected for thermal emission from the NS surface modulated by its rotation. A similar situation is observed, e.g. for the well-studied and also radio-quiet pulsar Geminga, where there is a broad single pulse of the thermal emission observed in the soft X-ray band accompanied by two sharp peaks of non-thermal emission seen in hard X-rays and gamma-rays \citep{mori2014}.

We calculated the X-ray pulsed fraction (PF) as $(I_{\rm max}-I_{\rm min})/(I_{\rm max}+I_{\rm min})$, where $I_{\rm max}$ and $I_{\rm min}$ are maximum and minimum intensities of the pulse profile. The intrinsic PF in the soft band, corrected for the background contribution, is $23\pm 6$ per cent. In the hard band, the data allow us to place only a $3\sigma$ (99.7 per cent) upper limit $\rm{PF}< 30$ per cent.

\section{Spectra of \j0633\ and its PWN}
\label{sec:j0633spec}

We extracted time-integrated spectra of the \j0633\ pulsar from the MOS and pn data using a circular aperture with the radius of 15 arcsec centred at the pulsar position, as measured by \chan. The PWN spectra were extracted from a solid region shown in the right-hand panel of Fig.~\ref{fig:chandra}, wherein circular regions around \j0633\ and the source `A', with radii of 20 and 15 arcsec, respectively, were excluded. All background spectra were extracted from a dashed region which is also shown in the right-hand panel of Fig.~\ref{fig:chandra}. Redistribution matrix (RMF) and ancillary response (ARF) files were created by the {\sc rmfgen} and {\sc arfgen} commands. The total number of counts extracted from the pulsar aperture in the 0.3--10 keV range is 2894, with about 2626 counts left after subtraction of the background. The total number of the PWN counts is 4884 with about 2504 counts being from the PWN itself.

To describe the non-thermal emission of the pulsar magnetosphere, we applied a PL model and we used another PL for the PWN emission. For thermal emission from the NS surface, we tried blackbody (\textit{bb}) and several models of the NS hydrogen atmosphere available in {\sc xspec}. Namely, we considered models \textit{nsa12} and \textit{nsa13}, which provide spectra of NSs with fully ionized  atmospheres and uniform radial magnetic fields $B=10^{12}$~G and $B=10^{13}$~G \citep{pavlov1995}. We also considered models describing NSs with partially ionized atmospheres, {\sc nsmax} \citep{ho2008}. Below, these  models are referred to by the same number codes as in {\sc xspec}. A model \textit{ns1260} is for the uniform radial field $B=4\times10^{12}$~G, which is close to the dipole field of \j0633\ estimated from the $\gamma$-ray timing. Models \textit{ns123100} and \textit{ns123190} are for the dipole magnetic field with $B=1.82\times10^{12}$~G at the magnetic pole. They differ by the angle between the magnetic dipole axis and the direction to the  observer, 0\degs\ and 90\degs, which is encoded by the last two digits of their numerical codes. Models \textit{ns130100} and \textit{ns130190} are the same but for larger $B=10^{13}$~G. In the dipole models, the NS temperature varies with the magnetic latitude due to the magnetic anisotropy of the heat transfer from the star interiors making the NS pole significantly hotter than the equator. For all thermal models, the gravitational redshift $1+z_g=[1-2.953 M_{\rm NS} (M_\odot)/ R_{\rm NS} ({\rm km})]^{-1/2}$ was fixed at 1.21, which corresponds to a reasonable NS with a mass $M_{\rm NS}=1.4 M_\odot$ and a circumferential radius $R_{\rm NS}=13$~km.

To describe the interstellar absorption, we applied an {\sc xspec} photoelectric absorption model {\sc phabs} with atomic cross-sections from \citet{bcmc1992ApJ} and solar abundances from \citet{anders1989GeCo}.

We performed spectral analysis in the 0.3--10 keV range simultaneously for \j0633\ and the PWN, assuming a common value of the absorption column density $N_{\rm H}$. We grouped spectra of \j0633\ and the PWN using the {\sc ftools grppha} command \citep{FTOOLS} with the condition that each spectral bin should contain at least one count. As a likelihood, we used the so-called \textit{W} statistic \citep{XspecManual}, which is the \textit{C} statistic \citep{cash1979} modified to account for Poisson background and which tends to $\chi^{2}$, for background-subtracted spectra, as the number of counts in each bin increases.

\begin{figure}
  \includegraphics[width=\columnwidth]{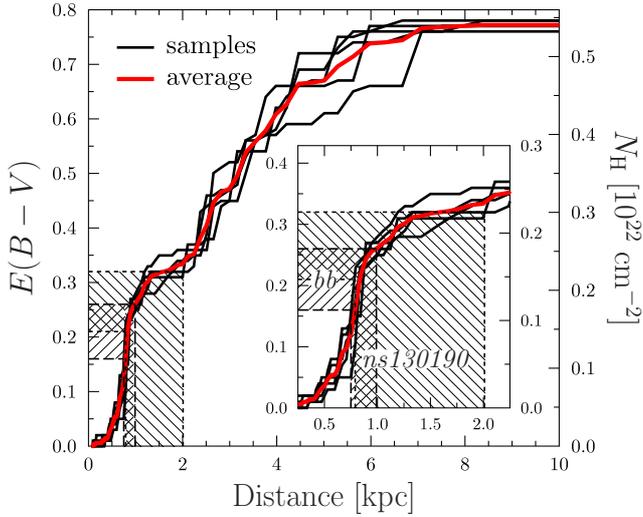}
  \caption{Relation between selective extinction $E(B-V)$ and distance in the direction towards the \j0633 pulsar, according to the 3D map of the dust distribution in the Galaxy presented by \citet{green2019}. A distance-extinction domain relevant to \j0633 is magnified in the inset. Hatched areas show 90 per cent credible intervals for the \textit{bb} and \textit{ns130190} atmosphere models. The right vertical axis represents values of absorption column density $N_{\rm H}$ obtained using an empirical relation between $E(B-V)$ and $N_{\rm H}$ proposed by \citet{Watson2011}.}
  \label{fig:extmaps}
\end{figure}

We analyzed spectra following the Bayesian approach and using an MCMC sampler proposed by \citet{goodman2010CAMCS} (GW). It is straightforward in the Bayesian inference to account for some additional information (\textit{prior}), which can help to better constrain the model parameters. In this respect, to constrain the radius of the thermally-emitting area on the NS surface $R$ and the distance to the NS $D$ separately, but not only their ratio, we used a relation between distance and the Galactic selective extinction $E(B-V)$ in the pulsar direction as a prior \citep[see e.g.][]{danilenko2015}. In Fig.~\ref{fig:extmaps}, we present such a relation obtained by means of a {\sc python} package {\sc dustmaps} \citep{green2018}. We produced it from a recent 3D map of the dust distribution in the Galaxy based on Gaia, Pan-STARRS 1 and 2MASS data \citep{green2019}. Five samples of the relation, shown in Fig.~\ref{fig:extmaps} by \old{gray}\aad{black} lines, are drown by means of the MCMC and represent  the corresponding posterior distribution \citep[see][for details]{green2019}. The thick \aad{red} line there is the median of the samples. To use the samples in X-ray data fitting, we transformed $E(B-V)$ to the absorption column density $N_{\rm H}$, the main parameter of the X-ray photoelectric absorption model. For the transformation, we applied an empirical relation $N_{\rm H}=(0.7\pm0.1)\times E(B-V)\times 10^{22}$~cm$^{-2}$ obtained by \citet{Watson2011} for the Galaxy using observations of X-ray afterglows of a large number of gamma-ray bursts.

To use all the samples shown in Fig.~\ref{fig:extmaps} in the MCMC, we do as follows. At each step of the MCMC, we randomly chose one of the samples and then, using the chosen sample, compute the distance from the current value of $N_{\rm H}$. In such an approach, only simulations of the distance actually depend on the additional information on the extinction. Other model parameters are simulated irrespective of such information. This approach is quite flexible as it allows one to account, in the same manner, for any number of additional relations, obtained from independent studies, between different model parameters.

\begin{figure}
  \includegraphics[width=\columnwidth]{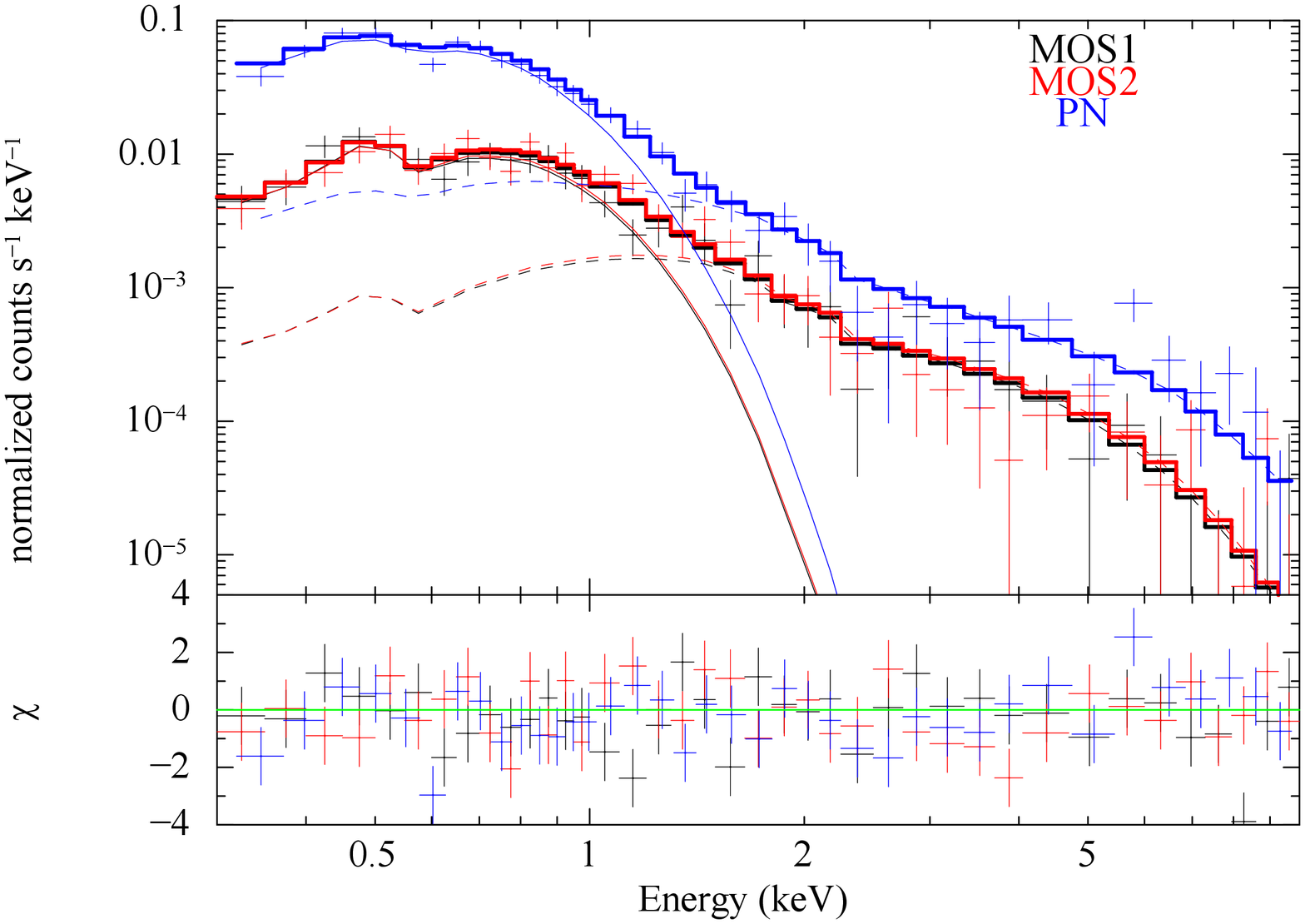}
  \includegraphics[width=\columnwidth]{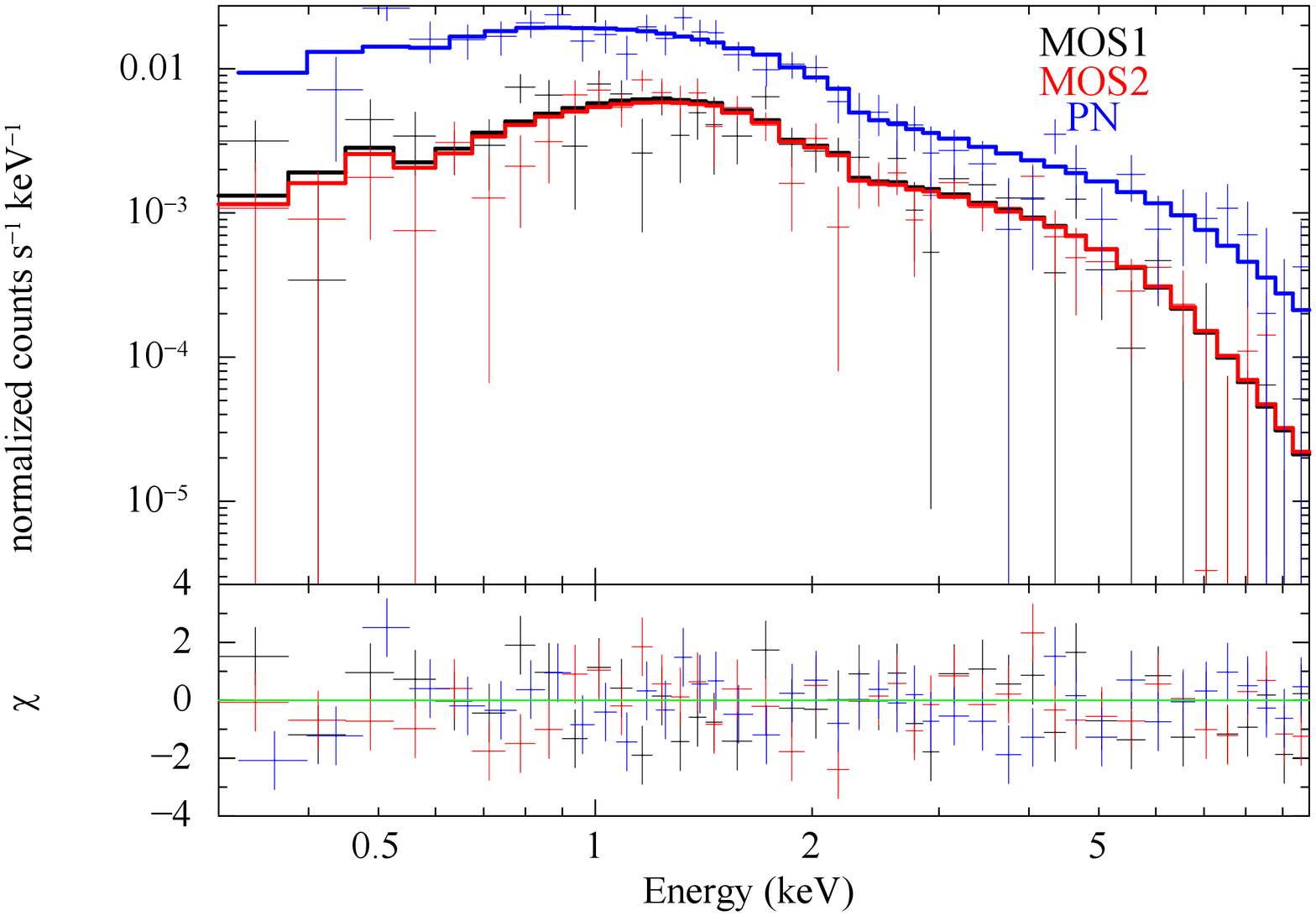}
  \caption{{\sl Top}: \xmm\ spectra of the \j0633 pulsar. The best-fit \textit{bb+PL} model (see text for details) is shown by the bold solid lines while the contributions of the \textit{bb} and \textit{PL} components are shown by thin solid and dashed lines, respectively. {\sl Bottom}: \xmm\ spectra of the \j0633\ PWN and the best-fit \textit{PL} model. The data from different instruments are shown by different colours. Spectra were grouped for illustrative purposes.}
  \label{fig:specpsr}
\end{figure}

In the GW sampler, we set a number of walkers, which is actually the only parameter of the sampler, being  $L=128$. Running $L$ walkers in the GW method can be compared to running $L$ independent \textit{one-sampler} methods, like the MH one, at a time. However, for at least some distributions, basically smooth and uni-modal, the former gives much smaller auto-correlation time $\tau$, a parameter which quantifies the simulation errors, than the MH and Gibs samplers do \citep{goodman2010CAMCS}. Note, that we used the MH algorithm instead of the GW one when searched for pulsations (Section~\ref{sec:timing}). Indeed, statistical models of periodic signals quite often have extremely multi-modal distributions. In such cases, the GW sampler, at least as it is, becomes much less efficient.

\begin{table*}
  \centering
  \caption{Best-fitting parameters of the pulsar and PWN spectral models.$^\dag$ 
  }
  \label{tab:x-fit}
  \renewcommand{\arraystretch}{1.4}
\begin{tabular}{cccccccccccc}
  \hline
Thermal           & $N_{\rm H}$         & \multicolumn{2}{c}{${T_{\rm s}^{\infty}}$}
& $R$
& $D$                 & $\Gamma_{\rm psr}$  & $K_{\rm psr}$       & $\Gamma_{\rm pwn}$  & $K_{\rm pwn}$       & $W$   & $\chi^{2}$ \\
model             & $10^{21}$ cm$^{-2}$ & eV              & $10^5\,$K            & km                  & kpc                 &                     & $10^{-5}$~keV$^{-1}$ &                    & $10^{-5}$~keV$^{-1}$ & \multicolumn{2}{c}{d.o.f.=} \\
                  &                     &                 &                      &                     &                     &                     & cm$^{-2}$~s$^{-1}$  &                     & cm$^{-2}$~s$^{-1}$  & \multicolumn{2}{c}{$769$} \\
\hline
 \multicolumn{12}{c}{A couple of hot spots on the surface:} \\
\textit{bb}       & $1.5^{+0.4}_{-0.3}$ & $120^{+8}_{-8}$ & $13.7^{+1.0}_{-0.9}$ & $0.8^{+0.5}_{-0.3}$ & $0.8^{+0.2}_{-0.1}$ & $1.9^{+0.3}_{-0.3}$ & $1.0^{+0.3}_{-0.3}$ & $1.6^{+0.1}_{-0.1}$ & $2.4^{+0.3}_{-0.3}$ & $778$ & $759$ \\
\multicolumn{12}{c}{The bulk of the NS surface:} \\
\textit{nsa12}    & $1.8^{+0.4}_{-0.3}$ & $51^{+5}_{-6}$  & $5.9^{+0.6}_{-0.7}$  & $7^{+9}_{-3}$       & $0.9^{+0.4}_{-0.1}$ & $1.6^{+0.3}_{-0.3}$ & $0.7^{+0.2}_{-0.2}$ & $1.7^{+0.1}_{-0.1}$ & $2.6^{+0.3}_{-0.3}$ & $779$ & $761$ \\
\textit{nsa13}    & $1.8^{+0.4}_{-0.3}$ & $53^{+5}_{-6}$  & $6.2^{+0.6}_{-0.7}$  & $6^{+8}_{-2}$       & $0.9^{+0.4}_{-0.1}$ & $1.5^{+0.3}_{-0.3}$ & $0.6^{+0.2}_{-0.2}$ & $1.7^{+0.1}_{-0.1}$ & $2.7^{+0.3}_{-0.3}$ & $778$ & $762$ \\
\textit{ns1260}   & $1.8^{+0.4}_{-0.3}$ & $53^{+7}_{-5}$  & $6.2^{+0.8}_{-0.5}$  & $6^{+9}_{-3}$       & $0.9^{+0.6}_{-0.1}$ & $1.3^{+0.4}_{-0.3}$ & $0.5^{+0.3}_{-0.1}$ & $1.7^{+0.1}_{-0.1}$ & $2.6^{+0.4}_{-0.3}$ & $781$ & $766$ \\
\textit{ns123190} & $1.8^{+0.4}_{-0.3}$ & $51^{+5}_{-6}$  & $5.9^{+0.6}_{-0.8}$  & $11^{+12}_{-6}$     & $0.9^{+0.6}_{-0.1}$ & $1.4^{+0.3}_{-0.3}$ & $0.5^{+0.3}_{-0.1}$ & $1.7^{+0.1}_{-0.1}$ & $2.7^{+0.3}_{-0.3}$ & $783$ & $769$ \\
\textit{ns130190} & $2.0^{+0.3}_{-0.5}$ & $54^{+5}_{-5}$  & $6.2^{+0.6}_{-0.6}$  & $10^{+17}_{-5}$     & $0.9^{+1.1}_{-0.1}$ & $1.3^{+0.3}_{-0.3}$ & $0.5^{+0.2}_{-0.2}$ & $1.7^{+0.1}_{-0.1}$ & $2.7^{+0.3}_{-0.4}$ & $782$ & $772$ \\
\multicolumn{11}{c}{Unrealistically large radius of the emitting area:} \\
\textit{ns123100} & $1.9^{+0.4}_{-0.3}$ & $28^{+4}_{-3}$  & $3.3^{+0.5}_{-0.3}$  & $50^{+96}_{-28}$    & $1.1^{+1.0}_{-0.3}$ & $1.5^{+0.3}_{-0.4}$ & $0.6^{+0.3}_{-0.2}$ & $1.7^{+0.1}_{-0.1}$ & $2.7^{+0.4}_{-0.3}$ & $781$ & $757$ \\
\textit{ns130100} & $2.2^{+0.4}_{-0.3}$ & $32^{+3}_{-3}$  & $3.7^{+0.4}_{-0.4}$  & $71^{+162}_{-31}$   & $1.2^{+1.4}_{-0.2}$ & $1.4^{+0.3}_{-0.4}$ & $0.5^{+0.3}_{-0.2}$ & $1.8^{+0.1}_{-0.1}$ & $2.8^{+0.4}_{-0.3}$ & $796$ & $775$ \\

\hline
\end{tabular}
\begin{tablenotes}
  \item $^{\dag}$ The parameters of each thermal model are the effective temperature $T_{\rm s}^\infty=T_{\rm s}/(1+z_g)$, as measured by a distant observer, and the circumferential radius $R$ of the NS thermally emitting area. The gravitational redshift $1+z_g$ is fixed at 1.21, when it matters. To distinguish between the two PLs, describing non-thermal emission of the pulsar and the PWN, their photon indexes $\Gamma$ and normalizations $K$ are marked by respective subscripts. A common equivalent hydrogen column density $N_{\rm H}$ is shared between the pulsar and the PWN spectra. \aad{Best-fitting values are maximal-probability estimates with errors corresponding to 90 per cent credible intervals; all values derived via the MCMC.}
\end{tablenotes}
\end{table*}

\begin{table*}
\centering
\caption{X-ray fluxes, luminosities and efficiencies of the pulsar and the PWN.$^{\dag}$}
  \label{tab:fluxes}
  \renewcommand{\arraystretch}{1.4}
\begin{tabular}{ccccccccc}
\hline
Thermal & Bol. flux & Bol. lum. & PSR flux & PWN flux & PSR lum. & PWN lum. & PSR eff. & PWN eff. \\
model   & $\log F_{\rm bol}^{\infty}$  & $\log L_{\rm bol}^{\infty}$ & $\log F^{\rm psr}_{\rm 2-10 keV}$ &  $\log F^{\rm pwn}_{\rm 2-10 keV}$ & $\log L^{\rm psr}_{\rm 2-10 keV}$ & $\log L^{\rm pwn}_{\rm 2-10 keV}$ &  $\log \eta^{\rm psr}_{\rm 2-10 keV}$ & $\log \eta^{\rm pwn}_{\rm 2-10 keV}$ \\
\hline
\multicolumn{9}{c}{A couple of hot spots on the surface:} \\
\textit{bb} & $-12.5^{+0.2}_{-0.1}$ & $31.4^{+0.3}_{-0.2}$ & $-13.48^{+0.07}_{-0.09}$ & $-12.93^{+0.04}_{-0.05}$ & $30.4^{+0.2}_{-0.1}$ & $31.0^{+0.1}_{-0.1}$ & $-4.6^{+0.2}_{-0.1}$ & $-4.1^{+0.1}_{-0.1}$ \\
\multicolumn{9}{c}{The bulk of the NS surface:} \\
\textit{nsa12} & $-12.2^{+0.2}_{-0.2}$ & $31.8^{+0.5}_{-0.3}$ & $-13.44^{+0.07}_{-0.09}$ & $-12.94^{+0.04}_{-0.06}$ & $30.5^{+0.3}_{-0.2}$ & $31.0^{+0.3}_{-0.1}$ & $-4.5^{+0.3}_{-0.2}$ & $-4.0^{+0.3}_{-0.1}$ \\
\textit{nsa13} & $-12.2^{+0.2}_{-0.2}$ & $31.8^{+0.6}_{-0.3}$ & $-13.44^{+0.08}_{-0.09}$ & $-12.94^{+0.04}_{-0.06}$ & $30.5^{+0.5}_{-0.2}$ & $31.0^{+0.4}_{-0.1}$ & $-4.5^{+0.5}_{-0.2}$ & $-4.0^{+0.4}_{-0.1}$ \\
\textit{ns1260} & $-12.1^{+0.2}_{-0.2}$ & $31.8^{+0.6}_{-0.3}$ & $-13.42^{+0.06}_{-0.10}$ & $-12.95^{+0.04}_{-0.05}$ & $30.6^{+0.4}_{-0.2}$ & $31.1^{+0.5}_{-0.1}$ & $-4.5^{+0.4}_{-0.2}$ & $-4.0^{+0.5}_{-0.1}$ \\
\textit{ns123190} & $-12.0^{+0.1}_{-0.2}$ & $32.2^{+0.5}_{-0.5}$ & $-13.44^{+0.09}_{-0.08}$ & $-12.95^{+0.04}_{-0.06}$ & $30.6^{+0.4}_{-0.1}$ & $31.0^{+0.4}_{-0.1}$ & $-4.5^{+0.4}_{-0.1}$ & $-4.1^{+0.4}_{-0.1}$ \\
\textit{ns130190} & $-12.0^{+0.2}_{-0.2}$ & $32.2^{+0.7}_{-0.5}$ & $-13.43^{+0.08}_{-0.09}$ & $-12.94^{+0.04}_{-0.06}$ & $30.6^{+0.4}_{-0.2}$ & $31.1^{+0.7}_{-0.1}$ & $-4.5^{+0.4}_{-0.2}$ & $-4.0^{+0.7}_{-0.1}$  \\
\multicolumn{9}{c}{Unrealistically large radius of the emitting area:} \\
\textit{ns123100} & $-11.7^{+0.3}_{-0.2}$ & $32.5^{+0.8}_{-0.5}$ & $-13.44^{+0.08}_{-0.09}$ & $-12.96^{+0.05}_{-0.04}$ & $30.7^{+0.6}_{-0.3}$ & $31.2^{+0.5}_{-0.3}$ & $-4.4^{+0.6}_{-0.3}$ & $-3.9^{+0.5}_{-0.3}$ \\
\textit{ns130100} & $-11.2^{+0.3}_{-0.2}$ & $33.1^{+0.8}_{-0.4}$ & $-13.44^{+0.09}_{-0.08}$ & $-12.97^{+0.05}_{-0.05}$ & $30.8^{+0.6}_{-0.2}$ & $31.3^{+0.6}_{-0.2}$ & $-4.3^{+0.6}_{-0.2}$ & $-3.8^{+0.6}_{-0.2}$ \\
\hline
\end{tabular}
\begin{tablenotes}
  \item $^{\dag}$ These are intrinsic, or unabsorbed, fluxes $F$ and luminosities $L$ of the pulsar and the PWN emission, derived using \aad{the same MCMC simulations as used to produce} Table~\ref{tab:x-fit}. For the thermal component, the bolometric fluxes and luminosities are given \aad{as seen by a distant observer}. For non-thermal PL components, we chose a range of 2--10 keV. Efficiencies $\eta$ of the pulsar and the PWN are ratios of the corresponding non-thermal luminosities to the total spin-down luminosity of the pulsar. Fluxes and luminosities are given in units of erg~s$^{-1}$~cm$^{-2}$ and erg~s$^{-1}$.
\end{tablenotes}
\end{table*}

To estimate $\tau$, we followed a method proposed by Dan Foreman-Mackey, along with a number of useful advises and instructive examples in {\sc python}.\footnote{\url{https://dfm.io/posts/autocorr/}} Convincing results are obtained when $\tau$ satisfies an empirical condition $\tau < N/50$, where $N$ is the total number of samples. To ensure that the condition was satisfied, we had to run GW walkers for about $T\approx 10^{5}$ times yielding $N=L\times T\approx 10^{7}$. We obtained $\tau \approx 10^{4} \ll N/50\approx 2\times 10^{5}$. This means that $\approx 10^{7}$ generated samples are equivalent to $\approx10^{3}$ independent ones, which is enough to provide a robust result.

Each of the considered spectral models has eight free parameters. In Table~\ref{tab:x-fit}, we show their maximal-probability-density values and equal-probability credible intervals computed from the MCMC simulations, as well as $W$, $\chi^2$ and the degrees of freedom (d.o.f) values demonstrating the fit qualities. For completeness, in Table~\ref{tab:fluxes}, we present corresponding thermal bolometric fluxes $F_{\rm bol}^{\infty}$ and luminosities $L_{\rm bol}^{\infty}$, as well as non-thermal fluxes ($F^{\rm psr}_{\rm 2-10 keV}$ and $F^{\rm pwn}_{\rm 2-10 keV}$) and luminosities ($L^{\rm psr}_{\rm 2-10 keV}$ and $L^{\rm pwn}_{\rm 2-10 keV}$) of the pulsar and PWN in the 2--10~keV range. The last two columns there also provide efficiencies of the transformation of the pulsar spin-down power to its own non-thermal emission and the emission of the PWN defined as $\eta^{\rm psr,pwn}_{\rm 2-10keV}=L^{\rm pwn,psr}_{\rm 2-10keV}/\dot{E}$. We also show the spectra of \j0633\ and its compact PWN in Fig.~\ref{fig:specpsr}, and an example of the best-fitting models.

Considering the $W$ and $\chi^2$ values from Table~\ref{tab:x-fit}, one can see that all the models are consistent with the data. The similar conclusion was drawn by \citet{danilenko2015} from the analysis of the \textit{Chandra} data, while they tried only \textit{bb} and \textit{ns1260} models for the thermal component\textbf{}. For these two models, our results are generally consistent with the results of that work, while the parameter uncertainties obtained here are much smaller. To our surprise, we found no evidence of the absorption spectral feature at 0.8 keV in the \textit{XMM-Newton} data, whose presence in the \textit{Chandra} data was claimed by \citet{danilenko2015}. This becomes clear from examination of the residuals of the spectral fit by a purely continuum spectral model presented in Fig.~\ref{fig:specpsr}. The presence of the feature in the \textit{Chandra} data and its absence in the \textit{XMM-Newton} data remains puzzling. It could be either a time variable feature, \yus{or a low count fluctuation, or} an unknown \textit{Chandra} instrument artefact.

It is remarkable that all the models suggest credible intervals for the distance and $N_{\rm H}$ that consistent with each other within uncertainties. This is illustrated by hatched areas in Fig.~\ref{fig:extmaps} and implies a conservative range of the distance to \j0633 being 0.7--2~kpc. The parameters of the pulsar and PWN non-thermal emission do not seem to depend significantly on the thermal model type either. At the same time, there is a predictably noticeable dependence of the inferred  thermally emitting area and temperature of the NS on the chosen thermal model.

The \textit{bb} model implies that the thermal component comes from a hot spot with a temperature of about 120 eV and a radius of about 0.8 km. The latter is about twice as large as $\sim 0.4$ km, the `classical' size of a pulsar hot polar cap heated by relativistic particles from the pulsar magnetosphere, estimated for \j0633\ by \citet{danilenko2015}. This discrepancy will be reconciled if we assume that both polar caps are seen simultaneously, due to the gravitational bending of light. However, in that case, we would see a double-peaked pulse profile in X-rays in contrast to what is actually observed (Fig.~\ref{fig:xmmvsfermi}). Alternatively, if the magnetic dipole of the NS is shifted in such a way that both polar caps occupy the same longitude we will still see a single pulse profile. Another alternative is that we see just one hot spot on the surface but of unusually large size, which could be caused, for example, by deviation of the surface magnetic field from the dipole. On the other hand, models \textit{ns123100} and \textit{ns130100}, assuming the NS magnetic axis directed to the observer, yield enormously large emitting area radii of about 50--70~km. Since the expected NS radius is in the range of 10--15 km, as predicted by various theoretical models \citep[e.g.][]{LattPrak2016} and confirmed by both electromagnetic \citep[e.g.][]{DegSul2018} and gravitational-wave \citep[e.g.][]{GWtoMR2018} observations, these models can be rejected.

The rest of the spectral fits, that is, those by atmosphere models \textit{nsa12}, \textit{nsa13} and \textit{ns1260}, describing NSs with the radial magnetic field, and models \textit{ns123190} and \textit{ns130190}, which are for NSs with the dipole magnetic field and the magnetic axis being orthogonal to the line of sight, give similar estimates of the effective surface temperature $T_{\rm s}^\infty \sim (5-7)\times 10^5\,$K and circumferential radii $R\sim 3-27$~km (see Table~\ref{tab:x-fit}). The credible intervals derived for radii are consistent with the expected radii of NSs. These models thus imply that the thermal spectral component of \j0633\ comes from the bulk of the NS surface.

It is worth discussing why models considered in the above paragraph give similar parameters. \citet{ho2008} show that spectra of these models are very similar in the considered photon energy range (see their figs 12 and 14). This is partially due to the fact that X-ray emission of an NS with the dipole field is dominated by warmer regions around magnetic poles, where the magnetic field is almost radial. The rest of the surface is colder and virtually cannot be seen in X-rays. At the same time, variation of the magnetic field strength within a reasonable range of $10^{12}-10^{13}$~G does not affect fitting results significantly.

To conclude this part, the thermal emission of \j0633\ is coming from either hot polar caps or the entire NS surface. In the latter case, the spectral analysis suggests that the magnetic axis should stay almost orthogonal to the line of sight during the NS rotation. Consequently, the angle between \aad{the magnetic and rotational axes}\old{and the line of sight, or the viewing angle,} can be close to either 0\degs \yus{(nearly aligned rotator)} or 90\degs \yus{(nearly orthogonal rotator)}: the \aad{exact alignment} or orthogonality is excluded by the detection of X-ray pulsations. \yus{This can naturally explain the absence of the radio emission. The latter is believed to be strongly beamed along the magnetic axis and its narrow beam just misses the observer when this axis remains nearly orthogonal during the rotation of \j0633. The models of pulsar evolution predict both the alignment and counter-alignment of the magnetic and rotational axes but neither of these two possibilities has been convincingly justified by observations \citep[see e.g.][and references therein]{2017arzamass}. We cannot discern between the two cases either.} Phase-resolved spectral analysis would be useful to distinguish between the blackbody and atmospheric models. However, the number of obtained EPIC-pn counts is not large enough to produce high signal-to-noise ratio spectra for these purposes.

\section{Diffuse emission}
\label{sec:diffuse}

\begin{table}
\renewcommand{\arraystretch}{1.2}
\caption{Best-fitting spectral model$^\dag$ of the cosmic background.}
\label{tab:bkg_par}
\begin{center}
\begin{tabular}{lc}
\hline
Temperature $T_1$, keV                               & 0.1 (fixed) \\
Normalization$^\ddag$ $N_1$, cm$^{-5}$ arcmin$^{-2}$ & $3.4^{+0.8}_{-0.9}\times 10^{-6}$ \\
Temperature $T_2$, keV                               & $0.14^{+0.02}_{-0.02}$ \\
Normalization$^\ddag$ $N_2$, cm$^{-5}$ arcmin$^{-2}$ & $1.6^{+2.2}_{-0.7}\times 10^{-4}$ \\
Temperature $T_3$, keV                               & $0.61^{+0.17}_{-0.10}$ \\
Normalization$^\ddag$ $N_3$, cm$^{-5}$ arcmin$^{-2}$ & $2.0^{+0.8}_{-0.8}\times 10^{-6}$ \\
Photon index $\Gamma$                                & 1.46 (fixed) \\
PL normalization $K$,                                & $6.4^{+3.0}_{-3.0}\times 10^{-7}$ \\
ph s$^{-1}$ cm$^{-2}$ keV$^{-1}$ arcmin$^{-2}$       &  \\
Column density $N_{\rm H}$, cm$^{-2}$                & $6.5\times 10^{21}$ (fixed) \\
$\chi^2$/N$_{\rm bins}$                              & 190/187 \\
\hline
\end{tabular}
\end{center}
\begin{tablenotes}
  \item $^\dag${\sc mekal}+({\sc mekal}+{\sc mekal}+PL)$\times${\sc phabs}. Temperature $T_1$ and normalization $N_1$ are for LHB emission, $T_2$ and $N_2$ are for the Galactic halo, photon index $\Gamma$ and normalization $K$ are for cosmological sources and $T_3$ and $N_3$ are for the additional thermal component (see text). All errors are at 90 per cent confidence. N$_{\rm bins}$ is the number of spectral bins.
  \item $^\ddag$Normalization of the {\sc mekal} model $N = \frac{10^{-14}}{4\pi D^{2}_{\rm cm}}\int n_e n_{\rm H}dV$, where $n_e$ and $n_{\rm H}$ are the electron and hydrogen number densities, $V$ is the volume of the emitting region and $D_{\rm cm}$ is the distance in centimeters.
\end{tablenotes}
\end{table}

\begin{table*}
\renewcommand{\arraystretch}{1.2}
\caption{Best-fitting parameters for diffuse emission from regions 1--4.}
\label{tab:dif_par}
\begin{center}
\begin{tabular}{lcccc }
\hline
Region                                & 1                      & 2                      & 3                      & 4                     \\
\hline
Column density $N_{\rm H}$, 10$^{21}$ cm$^{-2}$ & $1.6^{+0.5}_{-0.4}$    & $2.3^{+0.7}_{-0.7}$    & $0.8^{+0.8}_{-0.6}$    & $2.1^{+1.2}_{-0.9}$   \\
Photon index $\Gamma$                 & $1.54^{+0.13}_{-0.12}$ & $1.50^{+0.15}_{-0.15}$ & $1.41^{+0.23}_{-0.21}$ & $1.56^{+0.26}_{-0.23}$\\
PL normalization $K$, 10$^{-5}$ ph s$^{-1}$ cm$^{-2}$ keV$^{-1}$ arcmin$^{-2}$                & $4.7^{+0.7}_{-0.6}$    & $6.8^{+1.4}_{-1.2}$    & $2.5^{+0.7}_{-0.6}$    & $2.8^{+1.0}_{-0.7}$   \\
Area, arcmin$^{2}$                   & 25                     & 14                     & 28                     & 95                    \\
$\chi^2$/$N_{\rm bins}$               & 211/194                & 114/114                & 157/145                & 264/239               \\
\hline
\end{tabular}
\end{center}
\begin{tablenotes}
  \item Errors are at 90 per cent confidence. $N_{\rm bins}$ is the number of spectral bins.
\end{tablenotes}
\end{table*}

\begin{figure}
\begin{minipage}[h]{1.0\linewidth}
\center{\includegraphics[width=0.65\linewidth,angle=-90,clip]{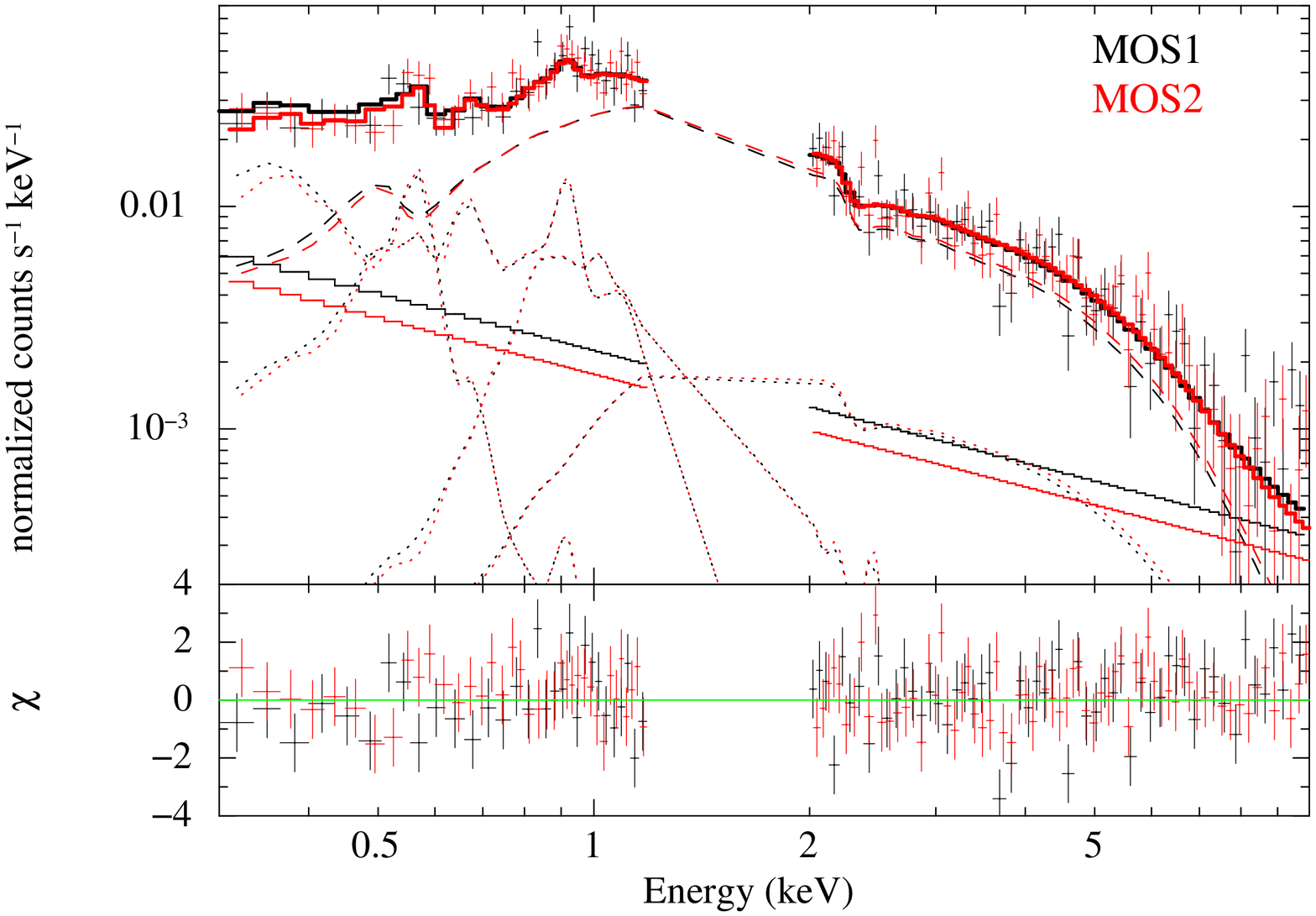}}
\end{minipage}
\begin{minipage}[h]{1.0\linewidth}
\center{\includegraphics[width=0.65\linewidth,angle=-90,clip]{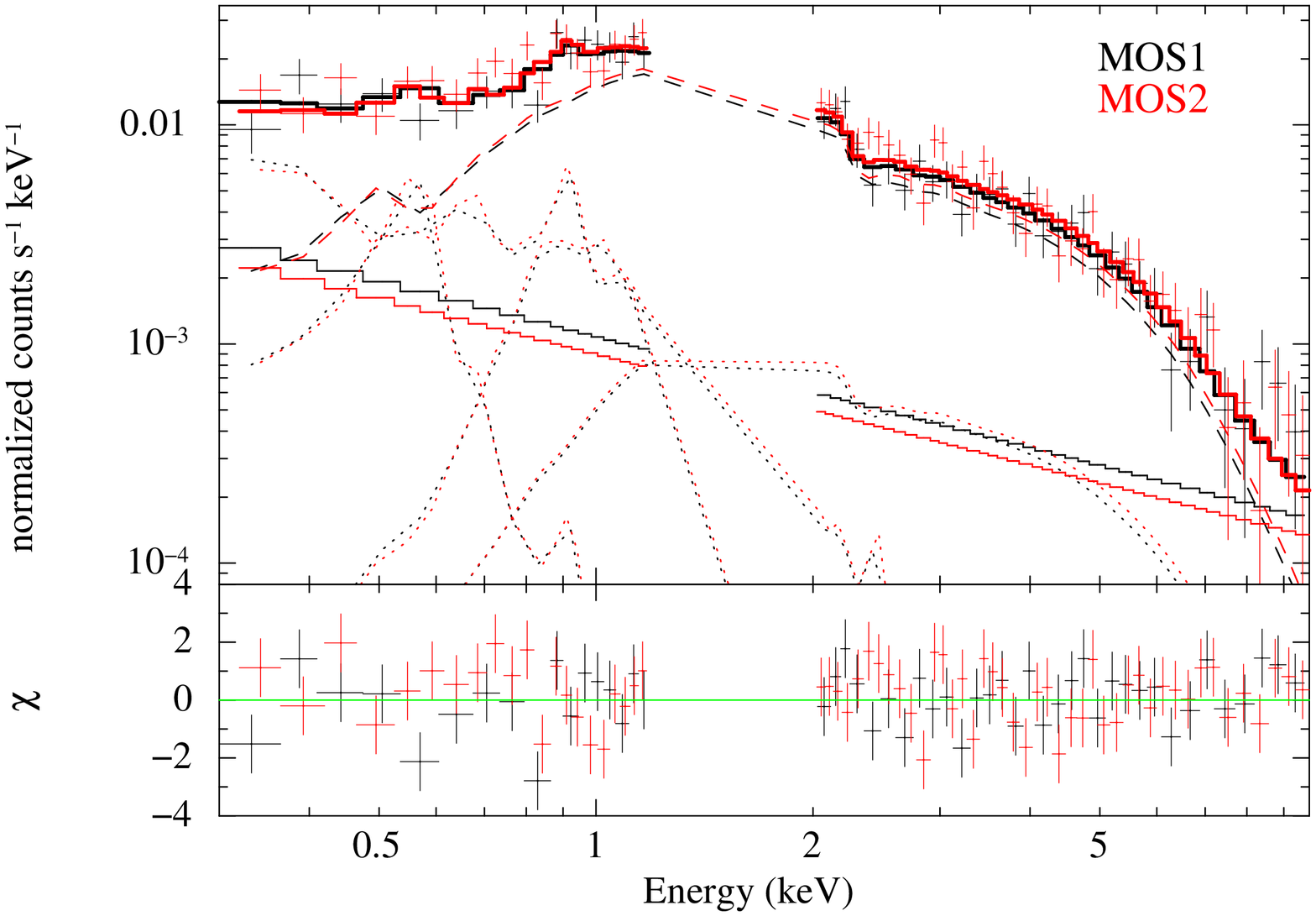}}
\end{minipage}
\caption{Observed spectra, best-fitting models and fit residuals for regions 1 ({\sl top}) and 2 ({\sl bottom}). The data from different instruments are shown by different colours. Solid bold lines show the best-fitting model (background + source); dotted lines show various components of the astrophysical background model; dashed lines are for PL components of the diffuse emission; finally, solid inclined lines show residual SP contamination.}
\label{fig:difspec}
\end{figure}

For the spectral analysis of the large-scale diffuse emission, we used the MOS data and the regions which are shown and numbered in the top right-hand panel of Fig.~\ref{fig:j0633comb}; the area inside the dashed box was used for the cosmic background. To extract the spectra and generate RMFs and ARFs, we applied an {\sc xmm-esas} task {\sc mos-spectra}. The spectrum of region 4 was obtained only from the MOS2 data since in the case of MOS1 it is partially projected onto the switched-off CCD. The {\sc mos\_back} tool was used to create model QPB spectra which were then subtracted from the data. All spectra were binned to ensure at least 50 counts per energy bin.

We fitted the spectra from the background region and regions 1--4 together in {\sc xspec} in the 0.3--10 keV energy range excluding 1.2--2 keV interval, which contains emission from the instrumental Al~K$\alpha$ and Si~K$\alpha$ lines. To account for the residual soft proton (SP) contamination, we utilized PL components convolved with diagonal response matrices. The photon indices of these components were assumed to lie in the range of 0.1--1.4 \citep[see][for details]{cookbook} and linked for the MOS1 and MOS2 data; for each detector, we tied together normalizations for different regions taking into account appropriate scale factors generated by the {\sc proton\_scale} task.

The spectra of diffuse emission were fitted with PL models. The cosmic background parameters were linked for all spectra assuming solid angle scale factors provided by the {\sc proton\_scale} tool for each region. A model of the cosmic background usually includes several components: еру unabsorbed thermal component with a temperature of about 0.1 keV from the Local Hot Bubble (LHB), the absorbed thermal component from the Galactic halo and the absorbed PL with $\Gamma$=1.46 from unresolved cosmological sources \citep{cookbook}. For the thermal components, we chose the optically thin plasma model {\sc mekal} \citep*{mewe1985}. The LHB temperature was fixed at 0.1 keV, and we take the value of the column density of 6.5$\times$10$^{21}$ cm$^{-2}$ based on the \citet{dickey1990ARAA} H{\sc i} maps, obtained using the {\sc heasarc} $N_{\rm H}$ toolю\footnote{\url{https://heasarc.gsfc.nasa.gov/cgi-bin/Tools/w3nh/w3nh.pl}} Additional components may be necessary since \j0633\ has a low Galactic latitude ($b=-0\fdg93$) \citep[see e.g.][and references therein]{ogrean2013}. We found that inclusion of one more thermal component in the model improves the fit. In this case $\chi^2$/d.o.f.~=~936/858 and the {\sc ftest} routine resulted in ф probability of chance improvement of $\approx7\times10^{-5}$; for the spectra of cosmic background itself this value is $\approx2\times10^{-3}$.

Best-fitting parameters for the astrophysical background are presented in Table~\ref{tab:bkg_par}, while parameters for the diffuse emission from regions 1--4 are given in Table~\ref{tab:dif_par}. Each of the spectra of regions 1--4 is well fitted by a single PL model, thus demonstrating the pure non-thermal nature of the extended emission with almost similar photon indices. Examples of the observed spectra for regions 1 and 2 together with best-fitting models are presented in Fig.~\ref{fig:difspec}. As seen, the fit residuals do not show any evidence of spectral features, confirming the above statement.

\section{Discussion}
\label{sec:discussion}

\subsection{\j0633 as a cooling NS}
\label{subs:j0633dis}

Assuming that the thermal emission of \j0633\ comes from the bulk of its surface, here we analyze the pulsar as a cooling NS. The respective analysis was done by \citet{danilenko2015} based on the \textit{Chandra} data obtained with a short exposure but it is worth a revision bearing in mind the better data quality obtained with \textit{XMM-Newton}. We take the pulsar characteristic age, $t_c = 59.2$~kyr, as a rough estimate of its true age $t$, and estimates of its surface temperature obtained from the X-ray spectral fits. We adopt errors of $\pm 0.3$, in $\log$-scale, to represent a realistic uncertainty of the true age.

As discussed in Section~\ref{sec:j0633spec}, the surface temperature estimate depends on the thermal model used in the spectral fit. We explore here the results for $T_s^\infty$ obtained utilising models \textit{ns123190} and \textit{ns130190}, which assume the magnetic dipole axis being perpendicular to the line of sight, since they
provide reasonable values of the circumferential radius.

\begin{figure*}
\centering
\includegraphics[width=0.9\textwidth]{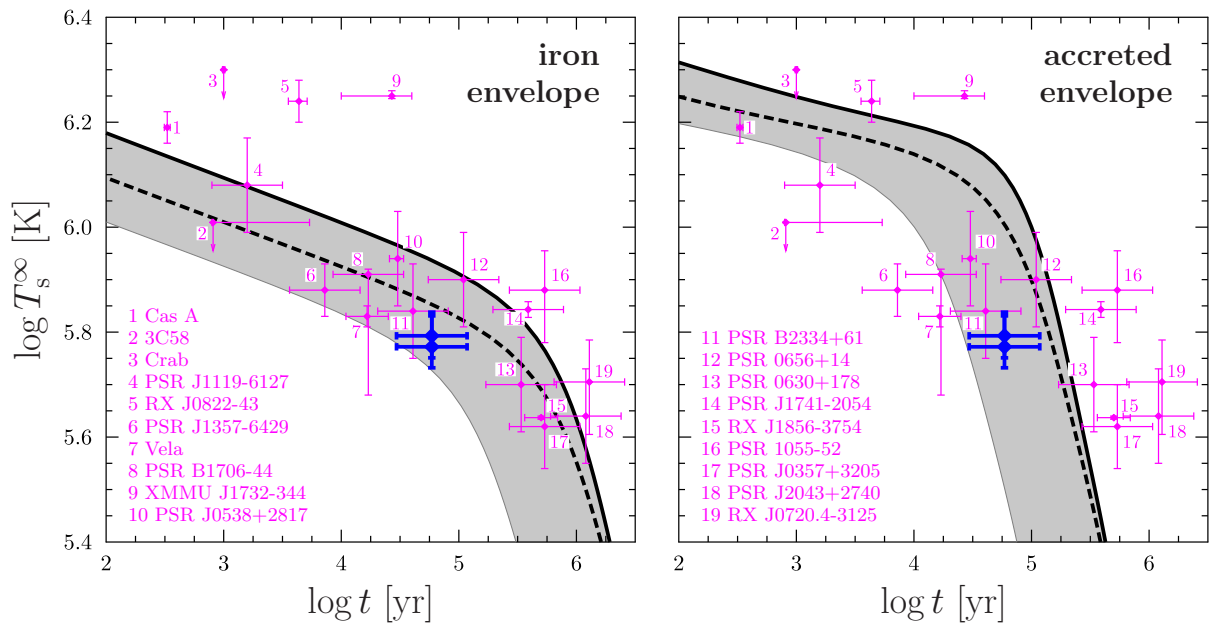}
\caption{\label{fig:cooling} \j0633 versus other cooling neutron stars and theoretical cooling scenarios for NSs with the iron ({\sl left}) and accreted ({\sl right}) heat blanket. The blue data points are the \textit{ns123190} and \textit{ns130190} atmospheric fits. Thick black curves show the standard cooling case (solid) and the cases with the neutrino luminosity 10 times (dashed) enhanced. The grey area approximately shows the domain that superfluid cooling NSs can occupy below the standard cooling curve. See the text for details.}
\end{figure*}

In Fig.~\ref{fig:cooling}, we compare \j0633 with a sample of other cooling NSs, $T_{\rm s}^\infty$ and $t$ of which are estimated from observations. The data are taken from table~1 of \cite{Bez2015}. While the sample does not include all the relevant observational data on cooling NSs obtained to date, it is quite representative for stars with magnetic fields not large enough to significantly affect cooling. The upper blue point shows the \j0633\ temperature obtained with the \textit{ns130190} model, which assumes a magnetic field at the pole of $10^{13}$~G. The lower one corresponds to the \textit{ns123190} model, which describes an NS with a field of $1.82\times 10^{12}$~G at the pole. Both cases are relevant to \j0633, since the spin-down dipole field estimate at the equator is $4.9\times 10^{12}\,$G. One can see that changing the magnetic field strength in reasonable limits does not significantly affect the position of \j0633 in the plot. It appears to be rather cold but not exceptional among other cooling NSs of a similar age.

For a general overview of the cooling theory, we refer to \citet{YakPeth2004} and \citet{PPP2015}. Briefly speaking, at $t\gtrsim 100$~yr, the NS interiors become almost isothermal (redshifted internal temperature is spatially constant), except for a thin outer layer, the heat blanketing envelope, whose properties define the relation between the surface and internal temperatures. The main cooling agents are the neutrino flow from the NS interiors (mainly from the core) and the photon flow from the surface, described by luminosities $L_\nu$ and $L_\gamma$, respectively. At $t\lesssim 10^5\,$yr, $L_\nu$ is much greater than $L_\gamma$ (neutrino-cooling stage), while for older NSs photon cooling is the most effective (photon-cooling stage). If $C$ is the total heat capacity of the star, its cooling rate depends on two ratios, $L_\nu/C$ and $L_\gamma/C$. $L_{\nu}$ and $C$ mainly depend on the equation of state of the core matter and pairing properties of baryons in the core. $L_\gamma$ depends on heat conducting properties of the outer envelope (heat blanket).\footnote{We do not discuss here the effects of strong internal magnetic field on cooling \citep[e.g.][]{PPP2015}, since \j0633 is not a magnetar.} Physics of the heat blanket is well-developed \citep[e.g.][]{PYCG2003} and generally parametrized by the surface magnetic field at the pole, $B_\text{pole}$, and the mass of matter accreted on the envelope, $\Delta M$. In the case of \j0633, $B_\text{pole} = 9.8\times 10^{12}\,$G (twice the equatorial field inferred from the spin-down), while $\Delta M$ is generally unknown and can vary from $\lesssim 10^{-17}\,$\Msun\ (iron envelope with lack of accreted material) to $\lesssim 10^{-7}\,$\Msun\ (fully accreted envelope). There are a lot of controversial models of the equation of state and pairing. Instead of specifying one, we adopt the model-independent approach \citep[e.g.][]{Yak2011}, which accounts for these phenomena.

The so-called standard cooling scenario \citep{Yak2011} is plotted in both panels of Fig.~\ref{fig:cooling} by solid black lines. In this scenario, the core is assumed to be nucleonic, with no baryon pairing and direct Urca process being prohibited. The main process responsible for neutrino emission is therefore the modified Urca (MUrca). Rapid increasing of the cooling curve slope (`the knee') at $t\sim 10^5$--$10^6\,$yr corresponds to transition from the neutrino-cooling to the photon-cooling stage. The luminosity $L_\nu$ and heat capacity $C$ are calculated according to \cite{Of2017} for an NS with $M=1.4\,$\Msun\ and $R=13\,$km, which is consistent with the redshift $1+z_g=1.21$ adopted in the spectral analysis. The left-hand panel corresponds to an NS with the iron envelope, $\Delta M = 10^{-17}\,$\Msun, while the right one is for $\Delta M = 10^{-7}\,$\Msun, i.e. a fully accreted heat blanket. Note that MUrca is treated here according to \cite{FM1979} whose approach has a lot of deficiencies \citep[see e.g.][for a comprehensive review]{SchSht2018}. The most important lack in their work is that it does not account for in-medium effects. The first effect to account for is the momentum dependence of the in-medium nucleon propagator \citep{SBH2018}. Ultimately this makes $L_\nu$ up to $\sim 10$ times larger than that of \cite{FM1979}. This case is shown by dashed lines in Fig.~\ref{fig:cooling}. There are several other in-medium effects (e.g. dressing of the virtual pion, see~\citealt{Vos2001}), but we do not  account for them here.

Nucleon pairing of different types (see \citealt{PageReview2014} for a review) affects NS cooling in a complex way. Pairing of any type ultimately reduces the heat capacity with a realistic lower limit of $~0.2\times\text{non-paired }C$. The neutron triplet pairing enhances the cooling rate $L_\nu/C$ up to a factor of $~100$ with respect to the standard cooling, due to the Cooper pairing neutrino emission.  These pairing effects are included in the minimal cooling paradigm \citep{GusMinCool2004,PageMinCool2004}. In Fig.~\ref{fig:cooling}, the grey-shaded strip is a schematic representation of a domain occupied by the minimal cooling curves.

Notice that this paradigm also suggests that there are cooling scenarios with $L_\nu/C$ less than the standard value. The corresponding cooling curves are located above the solid black lines in Fig.~\ref{fig:cooling}. These cases are not relevant to \j0633, and we do not consider them here.

One sees that the realistic fits of \j0633 spectra are consistent with various cooling scenarios for both accreted and iron heat blanket models. \j0633 can be either at the neutrino-cooling stage, with iron heat blanket and the cooling rate $L_\nu/C$ significantly enhanced due to superfluidity or in-medium effects, or at the photon-cooling stage, with accreted envelope and the heat capacity $C$ essentially suppressed by neutron pairing.

Several notes have to be made after this. First, remember that the magnetic field inferred from pulsar timing is just an estimate accurate up to a factor of a few \citep[e.g.][]{BAB2017}. Variation of the magnetic field within this uncertainty does not affect the surface temperature derived from the spectral fits. For instance, the two points marking the \j0633 position in Fig.~\ref{fig:cooling} correspond to atmosphere models whose magnetic fields differ by a factor of five, but their temperatures are consistent. We also studied how such field variations can affect \j0633 cooling, and found that this effect is insignificant. Second, we have not considered the direct Urca processes in the core (the so-called `rapid cooling'; e.g. \citealt{YakPeth2004}), nor have we allowed any hyperons. In the case of the iron heat blanket, a treatment of the fits of \j0633 in terms of hyperon or `rapid' cooling can be relevant but quite tricky \citep[e.g.][]{Raduta2018,Tolos2018}, and these scenarios are beyond the scope of this paper. Finally, it is not enough to just explain a given NS by some specific cooling scenario. The real challenge is to find a model that explains simultaneously all the set of cooling NSs shown in Fig.~\ref{fig:cooling}, but this is a much more complex task.

\subsection{Diffuse emission}
\label{subs:diffuse}

\begin{figure*}
\begin{minipage}[h]{0.495\linewidth}
\center{\includegraphics[width=0.90\linewidth,trim={0 0.0cm 0 0.cm},clip]{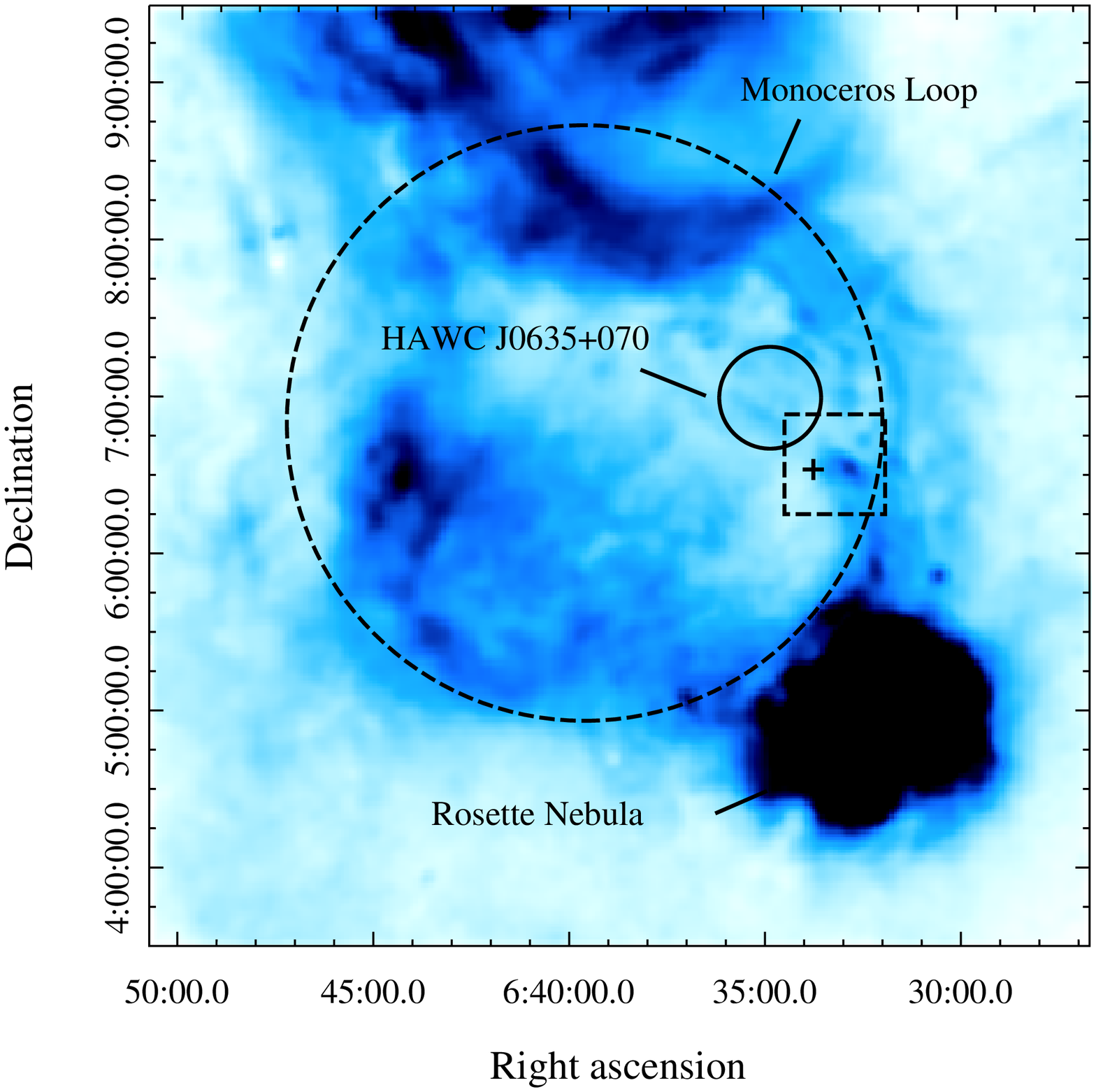}}
\end{minipage}
\begin{minipage}[h]{0.495\linewidth}
\center{\includegraphics[width=0.9\linewidth,trim={0 0.cm 0 0.cm},clip]{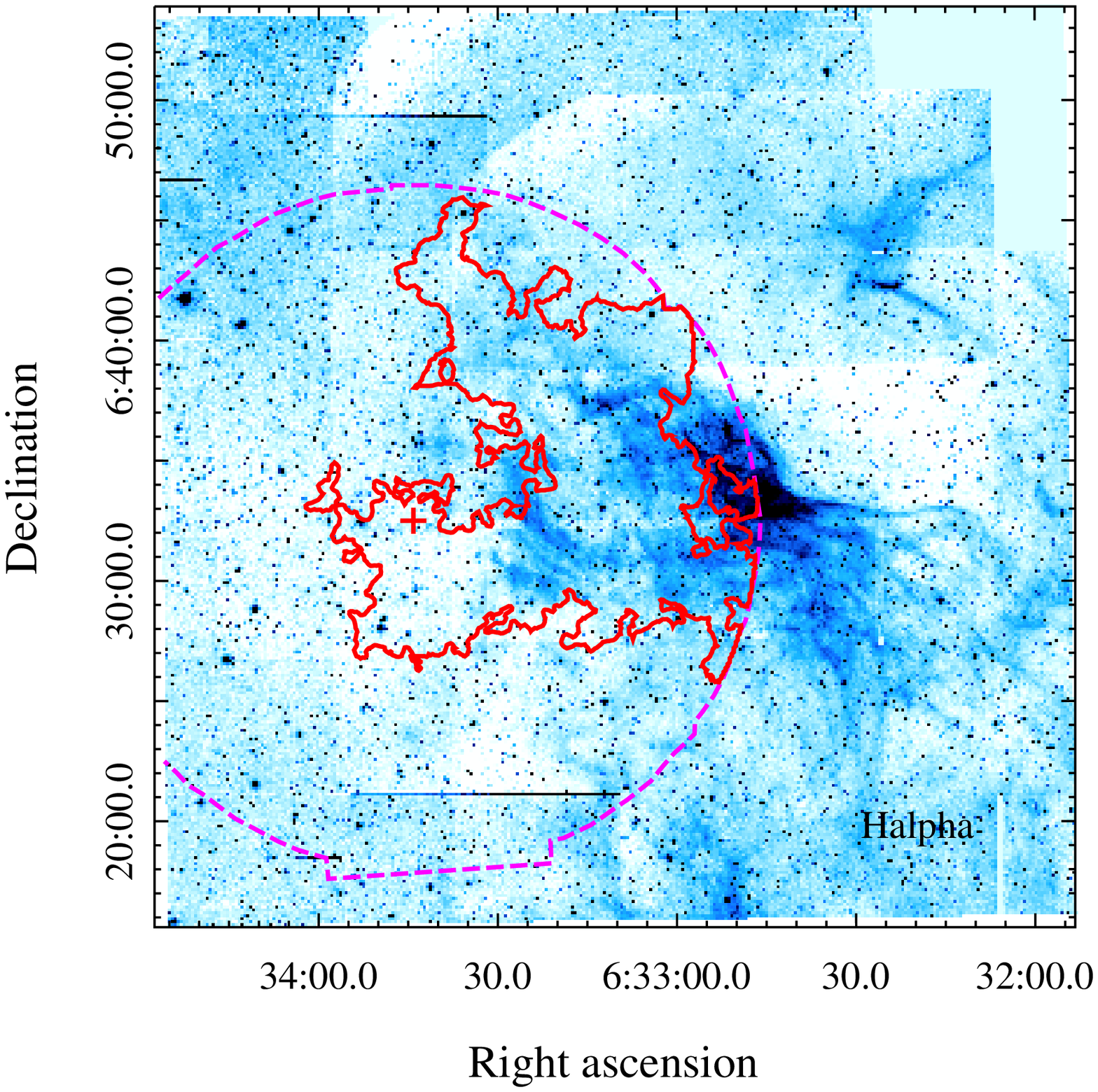}}
\end{minipage}
\begin{minipage}[h]{0.495\linewidth}
\center{\includegraphics[width=0.9\linewidth,trim={0 0.cm 0 0.cm},clip]{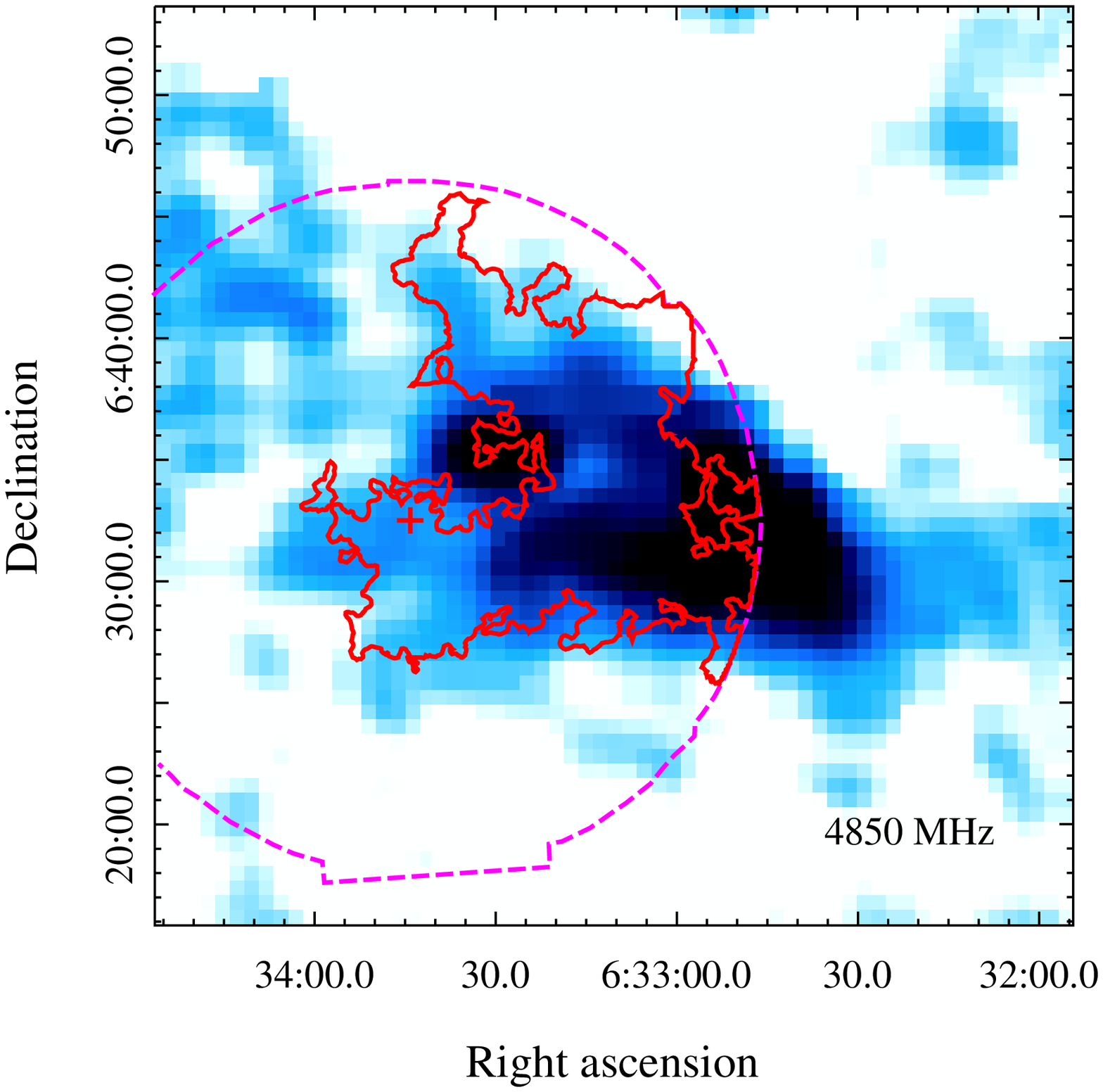}}
\end{minipage}
\begin{minipage}[h]{0.495\linewidth}
\center{\includegraphics[width=0.92\linewidth,trim={0 0.cm 0 0.cm},clip]{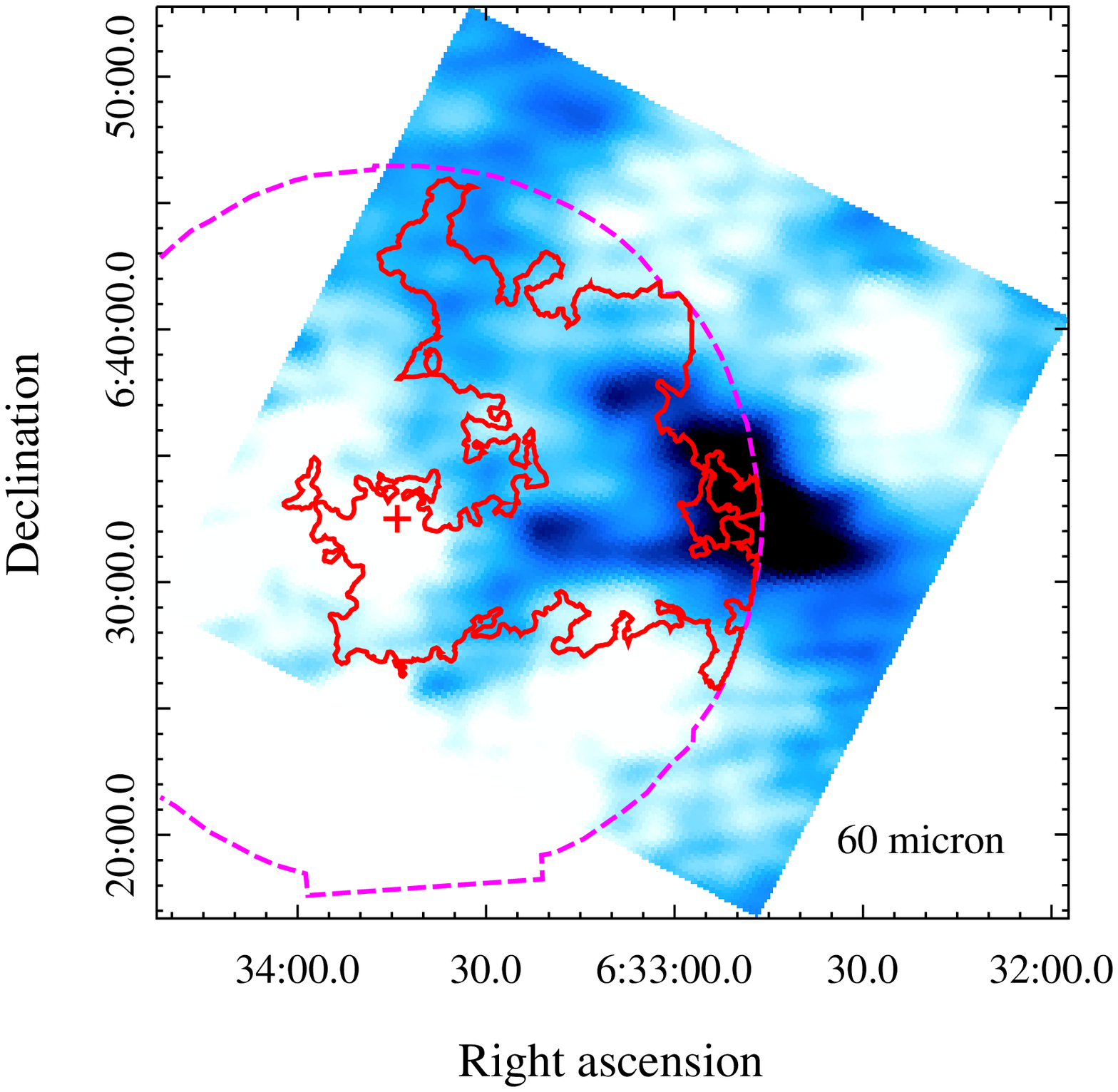}}
\end{minipage}
\caption{{\sl Top left}: H$\alpha$ image of the Monoceros loop region taken from SHASSA. The Monoceros loop SNR is marked by the dashed circle with a diameter of $\approx$4\degs. The solid circle indicates the position (RA=98\fdg71, Dec.=+7\fdg00) and a Gaussian 1$\sigma$ extent (0\fdg65) \citep{brisbos2018} of HAWC J0635+070. The Rosette nebula is also labelled. \avk{The dashed box with sizes of $\approx$38~$\times$~38 arcmin marks the region, enlarged in other panels, in which images obtained from different surveys are shown: IPHAS ({\sl top right}), 4.85 GHz Sky Survey ({\sl bottom left}) and the IRAS Galaxy Atlas ({\sl bottom right}). \xmm\ contours of diffuse emission are overlaid (contours correspond to intensity of 9 counts~s$^{-1}$~deg$^{-2}$ in the 0.4--7 keV band).} The \j0633\ position is shown by the `+' symbol. The \xmm\ combined MOS1+MOS2 FOV is indicated by the dashed line.}
\label{fig:bands}
\end{figure*}

The \j0633\ position is projected onto a large shell-type SNR Monoceros loop (G205.5+0.5). It was estimated to be 30--150 kyr old \citep{welsh2001}, which is in agreement with the \j0633\ characteristic age. The distance to the SNR is uncertain. Most estimates are about 1.6 kpc \citep[e.g.][]{borka2009}, which is compatible with the distance to the Rosette Nebula. It was assumed that these objects are interacting \citep[see][and references therein]{xiao2012}. However, two new estimates of the distance to the SNR appeared recently: \citet{zhao2018} obtained 1.98 kpc (and 1.55 kpc to the Rosette Nebula) while \citet{Yu2019} derived 0.94 or 1.26 kpc. These new estimates are apparently not consistent with the idea of  any interaction between the remnant and the nebula.

We checked whether the large-scale diffuse emission seen by \xmm\ has a thermal origin, i.e. it may be attributed to the Monoceros loop SNR. Using  the {\sc mekal} model instead of the PL resulted in temperatures of $\gtrsim 10$ keV for all the regions. They are too large  for the thermal emission of an evolved SNR such as the Monoceros loop. Therefore, the non-thermal origin of the emission seems to be more favourable.

Using various sky surveys, we found that the X-ray diffuse emission is projected on the edge of an extended clump detected in different bands: Effelsberg 11 and 21 cm radio continuum surveys of the Galactic plane \citep{furst1990,reich1997}, the Sino-German {$\lambda$}6 cm polarization survey of the Galactic plane\footnote{11, 21  and 6 cm images are available at \url{https://www3.mpifr-bonn.mpg.de/survey.html}} \citep{gao2010}, the 4.85 GHz Sky Survey\footnote{4.85 GHz images are available at \url{https://skyview.gsfc.nasa.gov/current/cgi/query.pl}.} \citep{condon1994}, the IRAS Galaxy Atlas\footnote{\url{https://irsa.ipac.caltech.edu/applications/IRAS/IGA/}} \citep[IGA; 60 and 100 $\mu$m,][]{cao1997}, the Southern H-Alpha Sky Survey Atlas\footnote{\url{http://amundsen.swarthmore.edu/SHASSA/}} \citep[SHASSA,][]{gaustad2001}, the INT Photometric H$\alpha$ Survey of the Northern Galactic Plane\footnote{\url{http://www.iphas.org/}} \citep[IPHAS; filters $r$ and H$\alpha$,][]{drew2005}, the Second Palomar Observatory Sky Survey\footnote{POSS-II data in digitized form are available at \url{http://stdatu.stsci.edu/cgi-bin/dss_form}.} \citep[POSS-II; red and blue plates,][]{lasker1996}, \textit{Spitzer} Galactic Legacy Infrared Midplane Survey Extraordinaire 360 \footnote{GLIMPSE data are available at \url{https://irsa.ipac.caltech.edu/data/SPITZER/GLIMPSE/}.} \citep[GLIMPSE360; 4.5 $\mu$m,][]{whitney2011}. In the case of the IPHAS data, stacked images were created using the {\sc casutools}\footnote{\url{http://casu.ast.cam.ac.uk/surveys-projects/software-release} {{\sc casutools}}} {\sc mosaic} command. The H$\alpha$ image of the Monoceros loop SNR is shown in the top left panel of Fig.~\ref{fig:bands}. The enlarged H$\alpha$ image of better spatial resolution, as well as 60-$\mu$m and 4.85-GHz images overlaid with contours of the extended emission revealed by \xmm, are shown in the top right-hand and bottom panels of Fig.~\ref{fig:bands}.

Optical emission of the clump can be produced by recombination lines of hydrogen, helium and carbon. The clump therefore may be a small dense cloud of interstellar matter. Thus, a likely origin of the emission from regions 3--4 is the interaction of particles accelerated in the shocks of the Monoceros loop with this cloud. The distance to the X-ray diffuse emission estimated using an $N_{\rm H}$--$D$ relation is compatible with the lower estimate of the distance to the SNR. The situation may be similar to that of SNR RX~J1713.7$-$3946, where the hard non-thermal X-ray features were assumed to be the result of interaction between dense molecular clumps and SNR shock waves \citep{sano2013}. SNR shock--cloud interactions can amplify magnetic field around clumps, which enhances X-ray emission around them \citep{inoue2012,sano2013}.

Extended emission in region 1 can be attributed to the fainter part of the PWN, i.e. shocked pulsar wind and shocked interstellar medium (ISM). PWNe spectra usually steepen with distance from a pulsar due to radiative losses of electrons. However, the \j0633\ PWN photon indices are in agreement, within uncertainties, with indices obtained for regions 1--4 (see Tables~\ref{tab:x-fit} and \ref{tab:dif_par}). The derived values are also in agreement with those obtained from \chan\ data by \citet{danilenko2015}, though their best-fitting indices are lower, that is, $\Gamma$~=~1.2--1.3, depending on the spectral model of the PSR+PWN system. \xmm\ has broad point spread function (PSF) wings, so the spectrum of the PWN in the pulsar vicinity may be somewhat softened by the pulsar emission contamination. Thus some steepening cannot be excluded though it is not enough to obtain typical values of photon indices in the case of synchrotron cooling ($\Gamma>2$).

There are some other PWNe where the same situation occurs. For instance, the photon indices of the tails of PSRs J1509$-$5850 and J0357+3205 do not show a significant dependence on distance from the pulsars, and the photon indices of the tails of PSRs B0355+54 and J1741$-$2054 shows only a hint of synchrotron cooling \citep[see e.g.][and references therein]{reynolds2017}. This may indicate an additional acceleration of particles within a tail. An alternative explanation is a high velocity of the outflowing matter and/or a low magnetic field \citep{reynolds2017}.

The elongated X-ray feature (region 2) may have various origins. \avk{Its orientation, almost transverse to the presumed pulsar proper motion, allows us to suggest that the feature may be an outflow misaligned from the pulsar as seen, for example, in the Lighthouse nebula \citep[see e.g.][]{pavan2016}. Alternatively, it can be explained in the same way as emission in regions 3--4, and that seems like a more favourable explanation because of a remarkable spatial coincidence between the X-ray emission seen in region 2 and the clump material revealed in various spectral bands, as it can be guessed from Fig.~\ref{fig:bands}.}

In Fig.~\ref{fig:bands}, we also indicate HAWC J0635+070, which was proposed as a TeV halo of J0633 \citep{brisbos2018}. However, its centre is shifted significantly from the pulsar position and thus its nature is still in question. It is interesting, that in the \xmm\ images (Fig.~\ref{fig:j0633comb}) there is some weak north-east protrusion, better seen in the hard 2--7 keV band, which directs to the TeV source and might indicate the association with it.

\begin{figure}
\begin{minipage}[h]{1.0\linewidth}
\center{\includegraphics[width=0.92\linewidth,trim={0 0.cm 0 0.cm},clip]{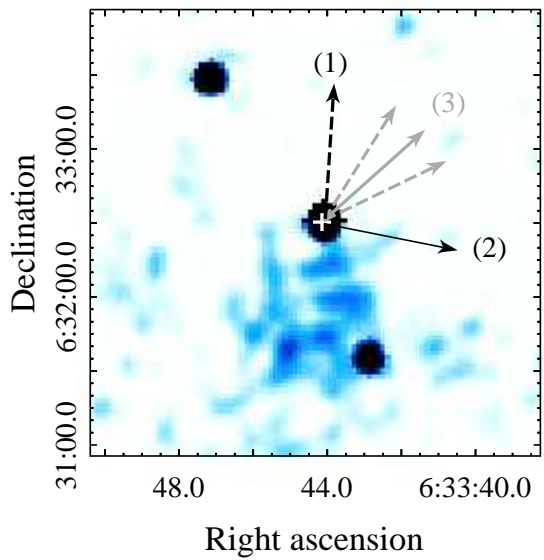}}
\end{minipage}
\caption{\avk{\chan\ image of the \j0633\ vicinity. The pulsar position is marked by the cross. The dashed black arrow (1), the same as the one shown in the left panel of Fig.~\ref{fig:chandra}, is the proper motion direction presumed from the PWN extension. The solid black arrow (2) shows the approximate direction in which the pulsar would be moving if it had been born near the centre of the Monoceros loop. Solid and dashed gray arrows (3) represent preliminary measurements of the pulsar proper motion, and its uncertainties, done by means of repeated \textit{Chandra} observations of \j0633 \citep{danilenko2019}.}}
\label{fig:pm}
\end{figure}

\avk{It was mentioned above that the characteristic age of \j0633\ and the Monoceros loop age are compatible. Nothing, therefore, would stop us from wondering if they are associated. If the pulsar was actually born somewhere near the centre of the Monoceros loop it would move approximately in the direction shown by the solid black arrow (2) in Fig.~\ref{fig:pm}, which does not follow the PWN extension at all. However, there are many examples of similar misalignment, e.g. the Lighthouse nebula mentioned above, and this would not therefore contradict the association. Meanwhile, we have repeated \chan\ observations of \j0633\ to measure its proper motion \citep{danilenko2019}. The pulsar's proper motion direction and
uncertainties are shown by solid and dashed gray arrows in Fig.~\ref{fig:pm}. It follows from these still-preliminary results that the pulsar is hardly associated with either the Monoceros loop or the Rosette Nebula. Fortunately, we found another possible birth site, an open stellar cluster Collinder 106 \citep{danilenko2019}.}

\section{Summary}
\label{sec:sum}

We analysed the \textit{XMM-Newton} observations of the \j0633 $\gamma$-ray pulsar. We confirmed previous investigations \citep{ray2011,danilenko2015} that the pulsar spectrum contains thermal and non-thermal components. The former can be equally well fitted by either the blackbody or magnetized neutron star atmosphere models. In the first case, the emission comes from hot spot(s), presumably pulsar polar caps, and, in the second case, it may originate from the entire NS surface. The derived spectral parameters of the pulsar and its PWN are in general agreement, within uncertainties, with those obtained from the \chan\ data \citep{danilenko2015}, uncertainties here being significantly smaller. However, new data do not confirm the absorption feature in the \j0633\ spectrum. Its apparent presence in the \textit{Chandra} data remains puzzling. It could be either a time-variable feature or an unknown instrument artefact. Using the interstellar extinction--distance relation, we better, comparing to previous studies, constrained the distance to the pulsar, 0.7--2~kpc.

We discovered X-ray pulsations from the pulsar. The pulse profile is broad and sinusoidal as expected for thermal emission modulated by NS rotation. The pulsed fraction in the 0.3--2 keV band is 23$\pm$6 per cent and its upper limit in the 2--10 keV range is $< 30$ per cent.

We analysed the cooling stage of the NS, accepting that the thermal emission is coming from the bulk of its surface with the effective temperature $T_{\rm s}^\infty \approx 6\times 10^5$~K, as shown by spectral fits. This result is quite insensitive to possible variations of the \j0633 magnetic field (within the reasonable range near the spin-down value). Depending on the cooling scenario, \j0633 can be either at the neutrino-cooling stage, with  the cooling rate significantly enhanced by nucleon super-fluidity or in-medium effects, or at the photon-cooling stage, if it has an accreted envelope and the heat capacity in its core essentially suppressed by neutron pairing.

Beside \j0633\ and its PWN, the \xmm\ observations revealed weak large-scale diffuse emission south, west and north-west of the pulsar. The part of this emission adjacent to the PWN may be attributed to the fainter emission of the shocked pulsar wind and shocked ISM. The brighter feature, elongated almost transverse to both the \old{presumed pulsar}\aad{PWN extent} and \aad{the preliminary direction of the pulsar proper motion measured  recently by \citet{danilenko2019}}, may be a misaligned outflow from the pulsar. The most favorable  explanation for other parts of the diffuse emission located at larger angular distances from the pulsar  is the interaction of particles accelerated in the shocks of the Monoceros loop SNR with the dense cloud of ISM detected in radio, IR and optical bands.

Deep X-ray observations with better spatial resolution are needed to carry out the detailed spatial and spatially-resolved spectral analysis of the large-scale diffuse emission. This may allow to separate the PWN emission from that caused by the interaction of the SNR shocks with a dense ISM. Time-resolved spectral analysis of different phases of the pulsar light curve obtained with better signal-to-noise ratio would allow one to establish, whether its thermal emission component comes from hot pulsar polar caps or from a cooler bulk of the NS surface. \old{The measurement of the \j0633\ proper motion may clarify the geometry of the system and the origin of the emission in different regions.}

\section*{Acknowledgments}
\aad{We would like to thank the anonymous referee for useful comments.} The scientific results reported in this article are based on observations obtained with \textit{XMM-Newton}, an ESA science mission with instruments and contributions directly funded by ESA Member States and NASA. For Fig.~\ref{fig:bands}, we used the data from the Southern H-Alpha Sky Survey Atlas (SHASSA), which is supported by the National Science Foundation. This paper also makes use of data obtained as part of the INT Photometric H$\alpha$ Survey of the Northern Galactic Plane (IPHAS) carried out at the Isaac Newton Telescope (INT). The INT is operated on the island of La Palma by the Isaac Newton Group in the Spanish Observatorio del Roque de los Muchachos of the Instituto de Astrofisica de Canarias. All IPHAS data are processed by the Cambridge Astronomical Survey Unit, at the Institute of Astronomy in Cambridge. The Second Palomar Observatory Sky Survey (POSS-II) was made by the California Institute of Technology with funds from the National Science Foundation, the National Geographic Society, the Sloan Foundation, the Samuel Oschin Foundation, and the Eastman Kodak Corporation. AD was supported by the Russian Foundation for Basic Research, grant 16-32-60129 mol\_a\_dk. The work of AK and DZ was supported by RF Presidential Programme MK$-$2566.2017.2 and the Russian Foundation for Basic Research, project 19-52-12013 NNIO\_a. The work of DO was supported in part by the Foundation for the Advancement of Theoretical Physics and Mathematics ``BASIS'' (Grant No. 17-15-509-1) and in part by the Russian Foundation for Basic Research, project 19-52-12013 NNIO\_a. The work of YAS was supported by the Russian Foundation for Basic Research, grants 16-29-13009 ofi\_m and 19-52-12013 NNIO\_a. DZ thanks Pirinem School of Theoretical Physics for hospitality.

\bibliographystyle{mnras}
\bibliography{refj0633}

\label{lastpage}

\end{document}